\documentclass[amsmath,trackchanges,twocolumn, tighten, times]{aastex702}

\usepackage{amsmath}
\usepackage{CJK}

\newcommand{\ha}{H$\alpha$}

\newcommand\hii{\ion{H}{2} }

\newcommand\oi{[\ion{O}{1}]}
\newcommand\oii{[\ion{O}{2}]}
\newcommand\oiii{[\ion{O}{3}]}

\newcommand\siii{[\ion{S}{3}]}
\newcommand\nii{[\ion{N}{2}]}
\newcommand\sii{[\ion{S}{2}]}

\newcommand\febpt{[\ion{Fe}{2}]$\lambda$12567/P$\beta$}
\newcommand\feii{[\ion{Fe}{2}]$\lambda$12567}
\newcommand\siiibpt{\siii$\lambda$9533/P$\gamma$}
\newcommand\heibpt{HeI$\lambda$10833/P$\gamma$}
\newcommand{\cue}{\textsc{Cue} }
\newcommand{\sigsfr}{$\Sigma_{\textrm{SFR}}$}
\newcommand{\fx}{$f_{\textrm{excess}}$}

\usepackage[dvipsnames]{xcolor}
\usepackage{enumitem}
\usepackage{hyperref}
\usepackage{CJK}

\begin{document}
\begin{CJK*}{UTF8}{gbsn}
\title{CECILIA: Multi-line Constraints on Excess Nebular Emission from Low-Ionization Gas at Cosmic Noon}

\author[0000-0002-6034-082X]{Caroline von Raesfeld}
\affiliation{Department of Physics and Astronomy, Northwestern University, 2145 Sheridan Road, Evanston, IL, 60208, USA}
\affiliation{Center for Interdisciplinary Exploration and Research in Astrophysics (CIERA), Northwestern University, 1800 Sherman Avenue, Evanston, IL, 60201, USA}
\email[show]{carolinevr@u.northwestern.edu}

\author[0000-0001-6369-1636]{Allison L. Strom}
\affiliation{Department of Physics and Astronomy, Northwestern University, 2145 Sheridan Road, Evanston, IL, 60208, USA}
\affiliation{Center for Interdisciplinary Exploration and Research in Astrophysics (CIERA), Northwestern University, 1800 Sherman Avenue, Evanston, IL, 60201, USA}
\email{allison.strom@northwestern.edu} 

\author{Audrey Clarendon} 
\affiliation{Department of Physics and Astronomy, Northwestern University, 2145 Sheridan Road, Evanston, IL, 60208, USA}
\affiliation{Center for Interdisciplinary Exploration and Research in Astrophysics (CIERA), Northwestern University, 1800 Sherman Avenue, Evanston, IL, 60201, USA}
\email{audreyclarendon2026@u.northwestern.edu}

\author[0000-0002-0361-8223]{Noah S. J. Rogers}
\affiliation{Center for Interdisciplinary Exploration and Research in Astrophysics (CIERA), Northwestern University, 1800 Sherman Avenue, Evanston, IL, 60201, USA}
\email{noah.rogers@northwestern.edu}

\author[0000-0002-0682-3310]{Yijia Li (李轶佳)}
\affiliation{Center for Interdisciplinary Exploration and Research in Astrophysics (CIERA), Northwestern University, 1800 Sherman Avenue, Evanston, IL, 60201, USA}
\email{yijia.li@northwestern.edu} 

\author[0000-0003-2385-9240]{Nathalie A. Korhonen Cuestas}
\affiliation{Department of Physics and Astronomy, Northwestern University, 2145 Sheridan Road, Evanston, IL, 60208, USA}
\affiliation{Center for Interdisciplinary Exploration and Research in Astrophysics (CIERA), Northwestern University, 1800 Sherman Avenue, Evanston, IL, 60201, USA}
\email{nathaliekorhonencuestas2029@u.northwestern.edu}

\author[0000-0002-8459-5413]{Gwen C. Rudie}
\affiliation{The Observatories of the Carnegie Institution for Sciences, 813 Santa Barbara Street, Pasadena, CA 91101, USA}
\email{gwen@carnegiescience.edu}

\author[0000-0002-6967-7322]{Ryan F. Trainor}
\affiliation{Department of Physics and Astronomy, Franklin \& Marshall College, 637 College Avenue, Lancaster, PA 17603, USA}
\affiliation{William H. Miller III Department of Physics and Astronomy, Johns Hopkins University, Baltimore, MD 21218, USA}
\email{ryan.trainor@fandm.edu}

\author[0009-0008-2226-5241]{Menelaos Raptis}
\affiliation{Department of Astrophysical Sciences, Princeton University, 4 Ivy Lane, Princeton, NJ 08544, USA}
\affiliation{Department of Physics and Astronomy, Franklin \& Marshall College, 637 College Avenue, Lancaster, PA 17603, USA}
\email{mr0194@princeton.edu} 

\author[0000-0002-1945-2299]{Zhuyun Zhuang}
\affiliation{Center for Interdisciplinary Exploration and Research in Astrophysics (CIERA), Northwestern University, 1800 Sherman Avenue, Evanston, IL, 60201, USA}
\email{zhuyun.zhuang@northwestern.edu}

\begin{abstract}
Spectroscopy from JWST has offered unique insights into elemental abundance patterns at Cosmic Noon ($z\sim2-3$), most notably a growing census of sub-solar S/O. These new measurements, coupled with new observations of [\ion{O}{1}]$\lambda6302$ at $z>1$, motivate an updated study of neutral and low-ionization emission within Cosmic Noon galaxies. In this letter, we present a new analysis of [\ion{O}{1}]$\lambda$6302 and [\ion{S}{2}]$\lambda\lambda$6718,33 in 19 galaxies observed using ultra-deep JWST/NIRSpec spectroscopy as part of the CECILIA survey. We find that it is necessary to consider sub-solar abundance patterns in sulfur at high-$z$ when analyzing galaxies on the S2-BPT, and that this correction reveals the potential of additional emission sources in CECILIA galaxies. Using the photoionization model emulator \textsc{Cue}, we find that $\sim48\%$ of observed [\ion{S}{2}] and $\sim55\%$ of observed [\ion{O}{1}] emission in CECILIA galaxies cannot be reproduced by conventional \ion{H}{2} region models. We find no clear evidence for turbulence or shocks driving this excess emission when comparing to existing models. Our findings imply that this excess emission may originate in diffuse ionized gas outside of \ion{H}{2} regions, though its nature may be different than that of low-$z$ DIG. Further, we find that including emission from gas outside of star-forming regions can incur a small bias up to 0.1 dex in 12+log(S/H).
\end{abstract}



\section{Introduction}
\label{sec:intro}

One of a star-forming galaxy's (SFG's) most information-rich signals is its emission-line spectrum. In SFGs, the strongest spectral lines originate in \hii regions photoionized by young, massive stars, where free electrons recombine with, or collisionally excite, the surrounding ions to produce recombination lines (RLs; e.g. H$\alpha$, H$\beta$) or collisionally-excited lines (CELs; e.g. \oiii, \nii, \sii, \oi, etc.). The strengths of these lines are sensitive to the complex physical conditions within the galaxy, most importantly the stellar radiation field ionizing the gas and the temperature ($T_e$) and density ($n_e$) of the interstellar medium (ISM) being ionized. These conditions in turn depend on the chemical abundances of the stellar population and ISM \citep{osterbrock2006}. Additional sources other than stellar photoionization can also produce line emission in star-forming galaxies, including active galactic nuceli (AGN) \citep{kewley2001agn}, shocks \citep{kewley2001, rich2010}, turbulence \citep{gray2017}, and diffuse ionized gas (DIG) \citep{madsen2006}. A galaxy's emission-line spectrum is therefore a complex creation governed by its ionizing sources, gas densities, temperatures, and chemical abundances.

This complexity, however, often makes it difficult to isolate how different processes drive emission-line production. Diagnostic diagrams such as the BPT and VO87 diagrams \citep{baldwin1981, veilleux1987} have leveraged the strong lines H$\beta$, \oiii$\lambda$5008, \oi$\lambda$6302, \nii$\lambda$6585, H$\alpha$, and \sii$\lambda\lambda$6718,33 in the local Universe to differentiate between ionizing sources of stellar populations, AGN, and shocks \citep{kewley2001, kauffmann2003}. As datasets have evolved, $z\sim0$ analyses have been able to additionally interrogate how a galaxy's complex ISM structure can alter its position on line-ratio diagrams. Two line transitions typically included in these diagrams, neutral \oi$\lambda6302$ and \sii$\lambda\lambda6718,33$ ($E_{\textrm{ion}}=10.36$ eV), can exist in partially-ionized gas outside of \hii regions, including in DIG and shock-heated gas. Studies of local spatially-resolved ($\sim1-2$ kpc) galaxies have shown that contribution from DIG outside of star-forming regions can elevate low-ionization CEL emission in a galaxy-integrated spectrum, shifting its location on these BPT diagrams towards higher \sii/\ha$\,$ and \oi/\ha$\,$ \citep{zhang2017, sanders2017}. Further, emission from shock-heated or turbulent gas can also shift a galaxy's location towards higher \nii/\ha, \sii/\ha, and \oi/\ha$\,$\citep{allen2008, ho2014, gray2017}.

As analyses of these same emission lines were extended to distant galaxies, the unique ISM properties at high $z$ began to be revealed. Numerous studies observed an offset between Cosmic Noon ($z\sim2-3$) and local galaxies on the N2-BPT \citep{erb2006metal,kewley2013, steidel2014, masters2014, shapley2015, strom2017, runco2022, sanders2023, shapley2025, schaerer2026}. The consensus view is that the offset is a natural consequence of a harder stellar ionizing spectrum at Cosmic Noon, and several studies at $z>2$ have measured a lower iron abundance (Fe/H) at fixed gas-phase metallicities (O/H) compared to local galaxies \citep{steidel2016, topping2020-composite, topping2020-individual,cullen2021, stanton2024-nirvandels}. 
This low Fe/H leads to a harder ionizing spectrum which imparts higher energies to the released electrons after ionization, thereby boosting CEL emission. Thus, it is necessary to consider both the detailed properties (e.g. density, multi-element chemistry) of the ISM along with the properties of the stellar population when interpreting emission-line spectra on diagnostic diagrams.

It is now well-established that the physical and chemical conditions of the ISM evolve with redshift: in addition to being ionized by harder radiation fields, it is also host to lower gas-phase metallicities \citep{erb2006metal, liu2008, sanders2015, theios2019, sanders2021, heintz2023, fujimoto2023, nakajima2023, curti2023, curti2024, morishita2024, sarkar2025, pollock2026, rogers2026, tang2026, stanton2026}, higher electron densities \citep{steidel2014, sanders2016, kaasinen2017, kashino2017, davies2021, isobe2023a, reddy2023, abdurrouf2024, topping2025-electrondensity, li2025-electrondensity}, and higher ionization parameters
\citep{sanders2016, sanders2023, cameron2023, nakajima2023, mascia2023, tang2025-ionization, topping2025-ionization, hayes2025, cleri2026}. 

Studies with JWST are additionally finding lower ISM abundances of sulfur and argon at fixed O/H \citep{rogers2024, stanton2025-excels, welch2024, foley2026, rogers2026,isobe2026-jades}. While oxygen is mainly produced in core-collapse supernovae (CCSNe), Fe, S, and Ar are produced in both CCSNe and Type 1a supernovae. The high O/Fe and low S/O and Ar/O abundance ratios therefore likely follow from the time delay between enrichment from CCSNe and Type Ia SNe \citep{kobayashi2009, kobayashi2020}, indicating that galaxies at Cosmic Noon are primarily CCSNe-enriched. 

The growing awareness of non-solar abundance patterns in sulfur at high-$z$, combined with new detections of \oi$\lambda6302$ at $z>1$ with JWST \citep{cameron2023, sanders2023, shapley2025, clarke2026, riviera-thorson2026}, motivate an updated analysis of neutral and low-ionization emission at Cosmic Noon where we can begin to constrain the complex ISM structure in star-forming galaxies.
In this letter, we characterize the \oi$\lambda6302$ and \sii$\lambda\lambda6718,33$ emission in a sample of 19 galaxies at Cosmic Noon, drawn from the CECILIA JWST/NIRSpec program. In Section \ref{sec:ceciliasamp}, we briefly describe CECILIA. In Section \ref{sec:bpts}, we present the N2-, S2-, and O1-BPT diagrams for the CECILIA sample and carry out a sample-level comparison to \textsc{Cloudy}  photoionization models \citep[v.23.01;][]{cloudy23}. In Section \ref{sec:Cue}, we model the line luminosities for individual CECILIA galaxies using the 
photoionization model emulator \textsc{Cue} \citep{li2025-cue} and quantify the excess low-ionization emission relative to the model predictions. In Section \ref{sec:emission}, we discuss potential sources of this excess observed emission. In Section \ref{sec:effects}, we quantify the potential impact of excess observed [S II]$\lambda\lambda6718,33$ on measured sulfur abundances. We summarize our conclusions in Section \ref{sec:conclusions}. 

Throughout this paper, we assume a $\Lambda$CDM cosmology with $H_0=70$ km/s/Mpc, $\Omega_\Lambda=0.7$, and $\Omega_m=0.3$. Solar abundances are adopted from \cite{asplund2021}: 12+log(O/H)$_\odot=8.69\pm0.04$, log(N/O)$_\odot=-0.86\pm0.08$, log(S/O)$_\odot=-1.57\pm0.05$. We refer to the position of spectral features using their vacuum wavelengths. We use S2 to refer to the ratio log(\sii$\lambda\lambda$6718,33/H$\alpha$), O1 to log(\oi$\lambda$6302/H$\alpha$), N2 to log(\nii$\lambda$6585/H$\alpha$), and O3 to log(\oiii$\lambda$5008/H$\beta$). When \oi$\,$ or \sii$\,$ is used, it refers to the line transitions \oi$\lambda$6302 and \sii$\lambda\lambda$6718,33. When discussing individual galaxies in the text, we use the galaxy ID and omit the "Q2343-" field name.

\section{CECILIA Sample}
\label{sec:ceciliasamp}
CECILIA is a Cycle 1 JWST GO program \citep[PID 2593;][]{strom2021} which uses electron temperatures and densities measured from faint rest-optical auroral lines to accurately measure the multi-element chemistry  in a sample of star-forming galaxies at $z\sim2.1-3.0$ \citep{rogers2026}. These galaxies were selected from the Q2343 field of the Keck Baryonic Structure Survey  \citep[KBSS;][]{steidel2010, rudie2012, strom2017} to span a representative range on the N2-BPT and the star-forming main sequence (SFMS; log(SFR) vs. log($M_\star$)). 

Observations with JWST/NIRSpec were conducted with two medium-resolution ($R\sim1000$) disperser-filter combinations: G235M/F170LP ($\lambda_{\textrm{obs}}=1.66-3.07\mu$m, $\lambda_{\textrm{rest}}\approx 4882-9323$\AA) and G395M/F290LP ($\lambda_{\textrm{obs}}=2.87-5.10\mu$m, $\lambda_{\textrm{rest}} \approx8441-15500 \mathrm{\AA}$). A total of 29.5 hours was obtained in G235M/F170LP along with 1.1 hours in G395M/F290LP. 

As part of KBSS, CECILIA galaxies have ancillary data from Keck/MOSFIRE in the $J-, H-$, and $K-$bands ($\lambda_\textrm{rest} \approx3000-7000$\AA, $R\sim3300-3700$) as well as archival photometry in $U_n$ through $K_s$, F140W, F160W, IRAC Ch1-4, MIPS 24 $\mu $m and narrow-band Ly$\alpha$ filters. Details about the parent KBSS sample and how the CECILIA galaxies were selected as well as information regarding the program design can be found in \cite{strom2023}.
The stellar masses and star-formation rates (SFRs), measured as described in \cite{korhonencuestas2025} and \cite{rogers2026}, span log$(M_{\star}/M_{\odot})= \;$8.3$-$10.5 with a median log$(M_{\star}/M_{\odot})=9.6$, and SFR$_{\textrm{SED}}=1.7-49.2$ M$_\odot$/yr with a median of 9.0 M$_\odot$/yr. The O/H abundances span 12+log(O/H)$=7.76-8.81$ dex, with an average of 8.29 dex \citep{rogers2026}.

\subsection{Spectroscopy and Emission Line Measurements}
The reduction of the Keck/MOSFIRE \citep{steidel2014, strom2017} and the JWST/NIRSpec \citep{rogers2026} data has been detailed in previous papers, along with the spectral energy distribution (SED) and emission-line fitting \citep{rogers2026}. Here, we review the details of the reddening correction and combination of the ground and space-based data.

As CECILIA galaxies have emission lines measured across Keck/MOSFIRE J, H, and K, and JWST/NIRSpec G235M/F170LP and G395M/F290LP, constructing a final emission-line flux catalog on a common scale requires cross-band and cross-instrument normalization. Further, a deviation from Case B recombination in the reddening-corrected Balmer lines from Q2343-D40 \citep{rogers2024} suggests a potential wavelength-dependent flux calibration error between G235M and G395M or an uncertain attenuation curve at high redshift \citep{sanders2025, reddy2026}. To mitigate these factors, we adopt a pseudo-reddening correction approach similar to that recommended by \cite{stasinska2025} and adopted by \cite{rogers2026}

We seek to place all emission lines on the scale of H$\alpha$ in NIRSpec G235M. For each galaxy, we use the highest-SNR detection of each line and take the ratio relative to the nearest resolved hydrogen line H$x$ (or P$x$ in the case of a Paschen line). 
Most lines require dust-correction when taking the ratio relative to the nearest H line; We dust-correct these ratios using the galaxy's E(B-V) calculated with the \cite{reddy2020} attenuation curve, the galaxy's Balmer decrement reported by \cite{rogers2026}, and a theoretical Balmer decrement of 2.82 in Case B conditions for $T_e$ = $1.25\times10^4$ K and $n_e$ = 300 cm$^{-3}$, using atomic data from \cite{storey1995}. As the ratios of \oiii$\lambda\lambda$4960,5008, \nii$\lambda\lambda$6550,85, He I 6678, \sii$\lambda\lambda$6718,33 and He I$\lambda$10833 with their respective neighboring H lines should have minimal dust attenuation, we calculate these ratios using uncorrected line fluxes.

Normalizing \oii$\lambda\lambda$3727,29 (detected only in MOSFIRE in this sample) is multi-step, as the closest significantly-detected hydrogen line (H$\beta$) is in a separate band. When H$\beta$ is available, \oii$\lambda\lambda$3727,29/H$\beta$ is corrected for reddening and differential slit losses. Three galaxies (BX341, BX350, and C31) lack a significant H$\beta$ detection from MOSFIRE, and we instead normalize \oii$\lambda\lambda$3727,29 to \oiii$\lambda$5008 (after correcting for reddening and slit losses), which appears in both MOSFIRE H-band (BX341, BX350) or K-band (C31) and in NIRSpec G235M. 

For \siii$\lambda\lambda$9071,9533, our approach depends on whether the lines are more significantly detected in G235M or G395M and which Paschen lines are detected in the same grating. If they fall in G235M, we normalize to Pa8 and do not reddening-correct the ratio. If they fall in G395M, where Pa8 is typically unresolved, we normalize to Pa9 (with no reddening correction) or Pa7 (with reddening correction) if Pa9 is unavailable.

We then multiply $I_\lambda$/H$x$ (or $I_\lambda/\textrm{P}x$) by the theoretical ratio H$x$/H$\alpha$ (or P$x/$H$\alpha$) to get $I_\lambda$/\ha. The theoretical H$x$/H$\alpha$ value is calculated with \textsc{pyneb}'s \citep{luridiana2015} \texttt{getEmissivity} function, using the individual galaxy's high-ionization zone $T_e$ and its $n_e$ (an average of $n_e$\oii$\,$ and $n_e$\sii, otherwise $n_e$\sii$\,$ when $n_e$\oii $\,$ is not available) if measured. When a value is not measured, we assume $T_e$ = $1.25\times10^4$ K and $n_e$ = 300 cm$^{-3}$,  reflective of galaxies at this redshift \citep{sanders2016, strom2018}.

We bring the line to the scale of NIRSpec G235M by multiplying this ratio $I_\lambda$/\ha$\,$ by the reddening-corrected line intensity of H$\alpha$ in G235M to get an absolute intensity $I_\lambda$. Finally, we calculate the line luminosity $L_\lambda=I_\lambda\times 4\pi D_L^2$, where $D_L$ is the luminosity distance determined from the spectroscopic redshift. This is the same approach used by \cite{rogers2026}, though now extended to all galaxies, not just those included in the $T_e$ sample. Three galaxies (BX391, D19, MD41) lack a detection of H$\alpha$ in NIRSpec, and are excluded from this analysis. 

\section{Emission-line Diagrams at Cosmic Noon}
\label{sec:bpts}



Emission-line studies of star-forming galaxies at  $z>1$ have grown significantly with JWST  \citep{cameron2023, sanders2023, topping2024, roberts-borsani2024, shapley2025, tang2026, clarke2026, stanton2026}, aiding in our characterization of star-forming (\ion{H}{2}) regions in these galaxies. Comparisons between photoionization models and these data on emission-line diagrams have begun to showcase how photoionization models with solar abundances are not entirely applicable at high redshift, with several recent analyses confirming the necessity of alpha-enhanced models at Cosmic Noon. With CECILIA, we seek to further interrogate the potential mismatch between Cosmic Noon galaxies and current photoionization modeling techniques, specifically in regards to the neutral and low-ionization species O$^0$ and S$^+$.


Both \oi$\lambda$6302 and \sii$\lambda\lambda$6718,33 were detected in the stack of CECILIA galaxies \citep{strom2023}; here, we report individual detections. The CECILIA sample has 18 detections of \oi$\,$ at a SNR $>3$ at a median SNR of 12, increasing the \oi$\,$ sample at $z\sim2-3$ by around $\sim25\%$. 
There are 22 detections of \sii, at a median SNR of 33. Motivated by these \oi$\,$ detections in the CECILIA and KBSS galaxies \citep{clarendon2025}, as well as the sub-solar S/O measured for the CECILIA galaxies \citep{rogers2026}, in Figure \ref{fig:bpt-pi} we plot the sample on the three common BPT diagrams, along with \textsc{Cloudy} photoionization models. We include galaxies from SDSS DR7 \citep{abazajian2009} as reference, filtering to only include galaxies with $0.04 \leq z \leq 0.1$ and H$\alpha$ measurements with SNR $>50$. We also plot the optical emission-line ratios of a stack of 30 KBSS galaxies (KBSS-LM1) at $z=2.40\pm0.11$ \citep{steidel2016} in pale yellow for comparison. We overplot \textsc{Cloudy} photoionization models (dark green manifolds) for comparison with the data,  as a first analysis of the S2-BPT with consideration of sub-solar S/O at high $z$.

\subsection{Photoionization Modeling with \textsc{Cloudy} for the CECILIA sample}
We compare our observations to photoionization models that consider the non-solar elemental abundances at Cosmic Noon. These models are an updated version of the \textsc{Cloudy} v.13 \citep{ferland2013} models which describe the full KBSS sample, used in \cite{steidel2016} and \cite{strom2017, strom2018}. The models are run using \textsc{Cloudy} version c23.01 \citep{cloudy23} with an input stellar ionizing spectrum from the stellar population synthesis code BPASS, version 2.2.1 \citep{eldridge2017}.
We adopt the default IMF with a slope of $-$2.35 over the range $0.5 \leq M_\star / M_{\odot} \leq 300$ assuming a constant star-formation history over 100 Myr.

We assume a plane-parallel geometry with a hydrogen density of 300 cm$^{-3}$, consistent with the electron densities from [O II] and [S II] \citep{rogers2026}. Further, we decouple the metallicity of the ionized gas $Z_{\textrm{neb}}$ and the stellar metallicity $Z_{\star}$ in our models, following previous high-$z$ photoionization modeling studies \citep[e.g.][]{steidel2016, strom2017, sanders2020}: high-$z$ galaxies are $\alpha$-enhanced, but without comprehensive $\alpha$-enhanced stellar population synthesis models this approach remains the most straightforward approximation.

\begin{figure*}
    \centering
    \includegraphics[width=0.98\textwidth]{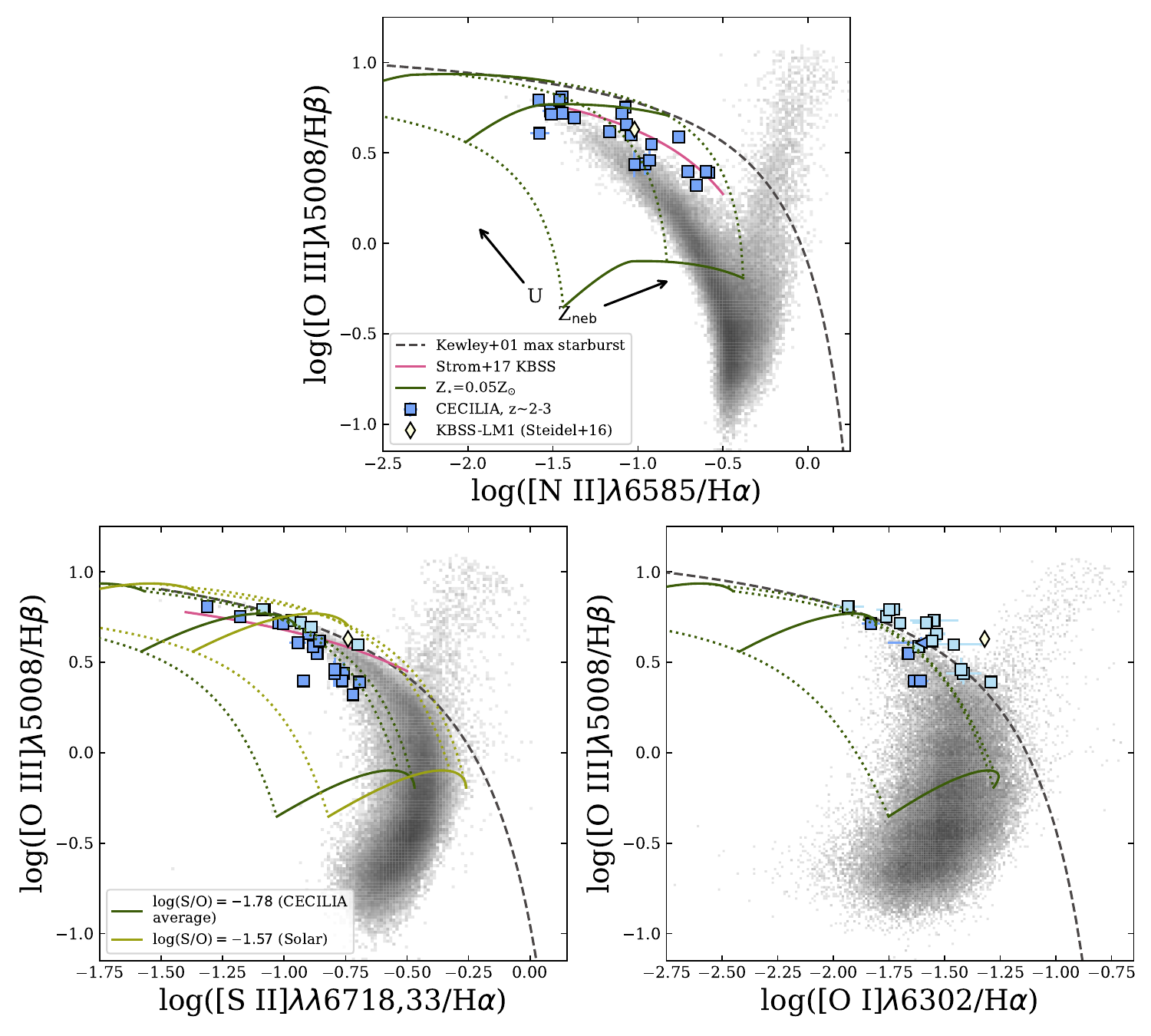}
    \caption{CECILIA galaxies on the N2-, S2-, and O1-BPT diagrams. We overplot the maximum starburst line from \protect{\cite{kewley2001}}, and include our fiducial photoionization models (dark green manifolds) with constant lines of log($U$)$ = [-3.5,-2,5, -1.5]$ (solid lines) and $Z_{\textrm{neb}}/Z_\odot = [0.1, 0.5, 0.9]$ (dotted lines). Upper limits are given as triangles. Photoionization models in the N2-BPT have been shifted to reflect the secondary increase in N/O at higher O/H (eq. \ref{eq-logno}). The fiducial model in the S2-BPT have been modified to match the average log(S/O) of the CECILIA sample, though we show the solar default in light green. While the CECILIA galaxies overlap with the photoionization models in the N2-BPT, several galaxies are above the models on both the S2- and O1-BPTs  (6 and 14, respectively; light blue points), indicating that there may be an additional source outside of star-forming regions contributing to this low-ionization and neutral emission.}
    \label{fig:bpt-pi}
\end{figure*}

Previous analysis of the FUV spectra of the KBSS-LM1 stack found best-fitting stellar metallicities of $Z_{\star}\sim0.001-0.002$, thus we make the assumption that CECILIA galaxies have similar stellar metallicities based on their similarity on the SFMS and their ionization conditions. 
While there is a range of stellar metallicities in the KBSS sample \citep{theios2019, strom2022} and other Cosmic Noon samples \citep{topping2020-individual, stanton2024-nirvandels}, we compare to the lowest representative stellar metallicity $Z_{\star}=0.001$ ($Z_{\star}=0.05Z_{\odot}$; dark green manifolds, Figure \ref{fig:bpt-pi}), as higher $Z_{\star}$ models are offset towards lower N2, S2, O1, and O3.
For gas-phase metallicity, we allow $Z_{\textrm{neb}}$ to vary from 0.01 to 2.0 times solar in steps of $\Delta$0.01. 

We parameterize the normalization of the input ionizing spectrum by the dimensionless ionization parameter (the ratio of the number density of H-ionizing photons to the number density of hydrogen, \textit{U=$\frac{n_\gamma}{n_H}$}), and allow log($U$) to vary between [$-$3.5, $-$1.5] in steps of $\Delta$0.01. In Figure \ref{fig:bpt-pi} we plot constant lines of log($U$) = [$-3.5, -2.5, -1.5$] as solid lines and constant lines of gas-phase metallicity $Z_{\textrm{neb}}/Z_\odot$ = [0.1, 0.5, 0.9] as dotted lines. 

We apply a metallicity-dependent scaling relation to the \nii$\,$ photoionization model fluxes that reflects the secondary increase in N/O with respect to O/H, according to 
\begin{equation}
\label{eq-logno}
    \textrm{log(N/O)} = (1.64\times \textrm{log($Z_{\textrm{neb}}$)}) - 0.86
\end{equation} 
 with a lower limit of log(N/O) $=-$1.5 \citep{strom2017}. 
 Though this relation may not be accurate for each individual galaxy, it offers a qualitative sample-level comparison for CECILIA. Applying other log(N/O)$-$log(O/H) relationships from high-$z$ literature that reach sufficiently low gas-phase metallicities for the CECILIA sample \citep{scholte2026} does not change our results.

An important update is the consideration of non-solar S/O. CECILIA galaxies have an average value of log(S/O) = $-$1.78$\pm$0.21 dex \citep{rogers2026}, 0.21 dex lower than the solar value of log(S/O)$_\odot=-1.57$. Previous photoionization model interpretations of the S2-BPT at high $z$ have assumed a solar S/O scaling, which may have obscured underlying physical conditions within galaxies. Similar to the post-processing scaling of \nii, we account for these non-solar S/O abundances by scaling the photoionization models in Figure \ref{fig:bpt-pi} from the default solar value to the average S/O in the CECILIA sample, subtracting 0.21 dex from the modeled log(\sii$\lambda\lambda$6718,33/H$\alpha$) ratio (the default solar value is given as the light green manifold). As we scale to the average in the sample, this may not entirely capture the appropriate abundance pattern for individual galaxies, but, similar to N/O, it offers a qualitative sense of how adjusting for sub-solar abundance patterns in sulfur may affect interpretations of galaxy properties based on BPT loci.

\subsection{Elevated O1 and S2 Line Ratios Relative to Photoionization Models}

While the CECILIA galaxies overlap with the fiducial photoionization model in the N2-BPT (top panel), the galaxies begin to deviate from the model surface in the S2-BPT (6 out of 22 galaxies, lighter blue, lower left panel), and the majority of the sample is above the model in the O1-BPT (14 out of 18  galaxies, lighter blue, lower right panel). Further, 11 out of 18 galaxies lie above the maximum starburst line of \cite{kewley2001} in O1-BPT, and 3 lie above it in the S2-BPT, further indicating there may be some additional emission contribution from a source such as shock-heated gas. 
Several high-redshift studies of star-forming galaxies have measured similarly elevated ratios of O1 and S2 compared to local \hii regions. \cite{sanders2023} find that CEERS composite spectra at four different redshift bins from $2.0 < z < 6.5$ are offset from $z=0$ \hii regions on the N2-, S2-, and O1-BPT, which the authors interpret as additional evidence of a harder ionizing spectrum at fixed O/H. \cite{shapley2025} find that $z>1.4$ AURORA galaxies are offset from local \hii regions in S2 and O1, similarly arguing for the existence of a harder ionizing spectrum. \cite{clarke2026} find that the locus of the 1.4 $< z <$ 7 JADES galaxies on the O1-BPT is elevated relative to local \hii regions and star-forming galaxies and suggest that there may be a strong contribution from supernova shocks in these galaxies, in addition to a harder ionizing spectrum.

  The observed deviations of CECILIA galaxies in the S2- and O1-BPTs from photoionization models, which account for the galaxies' harder ionizing spectra and non-solar abundance patterns, indicate that photoionization from massive stars may not be entirely responsible for the emission in these galaxies. To investigate the presence of an additional emission source in CECILIA galaxies, we must first constrain the amount of emission in each individual galaxy that is not described by photoionization modeling. 


\section{Predicting Line Emission from Photoionization with \textsc{Cue}}
\label{sec:Cue}

To assess how well current photoionization models can reproduce the line luminosities in individual CECILIA galaxies, we use the spectral analysis code \textsc{Cue} \citep{li2025-cue}. \textsc{Cue} is unique compared to other photoionization modeling tools in that it is agnostic to the input ionizing spectrum. The code takes in an observed emission-line dataset and reconstructs a 4-part piecewise power-law ionizing spectrum, an ionization parameter log($U$), as well as properties of the ionized gas (hydrogen number density $n_{\textrm{H}}$, gas-phase metallicity O/H, and abundance ratios N/O and C/O). These parameters are used to predict line luminosities using a neural net emulator of \textsc{Cloudy} \citep[version 22.00;][]{cloudy22} trained on a diverse set of ionizing spectra from star-forming galaxies to AGN.
The parameters are varied using the dynamic nested sampler \texttt{dynesty} \citep{speagle2020,kosposov2022}, and the likelihood is calculated using the difference between the predicted and observed line luminosities. In our \textsc{Cue} runs, we adopt the same priors as \cite{li2025-cue}; these are uniform distributions across the full parameter space used in the neural net training, and are given in Table 1 in \cite{korhonencuestas2026}.


We run \textsc{Cue} on all CECILIA galaxies with H$\alpha$ detected in the JWST/NIRSpec data. However, we must make several modifications. First, the photoionization models that \textsc{Cue} was trained on assume solar S/O and Ar/O. To mitigate bias from assuming solar S/O and Ar/O for galaxies with known sub-solar S/O and Ar/O, we exclude all sulfur and argon lines from the \textsc{Cue} fits \citep{korhonencuestas2026}. 
Further, we exclude low-ionization lines ($E_{\textrm{ion}} < 13.6$ eV), such as \oi$\,$ and \sii$\,$, to ensure that the galaxy parameters are fit only using lines that we are confident originate from \hii regions. We also exclude the He I$\lambda$10833 line. This line is sensitive to density \citep{izotov2014, aver2015, berg2026, skillman2026}, but due to neutral helium's high ionization potential, it can be produced in different components of the multi-phase ISM, a stratification we see in both low-$z$ nebulae \citep{mendez-delgado2023, rogers2026-yp} and high-$z$ galaxies \citep{martinez2025, topping2025-electrondensity}. \textsc{Cue} assumes a single hydrogen density in the modeling, and including this line can therefore obscure the density of the low-ionization species constrained by \oii$\lambda\lambda$3727,29 and \sii$\lambda\lambda$6718,33. Runs including this line returned densities $\sim$0.6 dex higher than the directly-measured electron densities from the \oii$\,$ and \sii$\,$ doublets, thus we exclude it in order to better understand the behavior of the low-ionization lines.

The CECILIA galaxy properties reported by \textsc{Cue} are generally in agreement with properties measuring using the direct method \citep{rogers2026}, including gas-phase O/H \citep{korhonencuestas2026}. We also find that the N/O inferred from \textsc{Cue} agrees with the $T_e$-based N/O. We input the full \textsc{Cue} posterior into the photoionization model emulator to generate model line luminosity distributions for each galaxy. 
We take the median of each line prediction distribution as the representative value and report the 16th and 84th percentiles to capture any non-Gaussianity.


\begin{figure}
    \centering
    \includegraphics[width=1\linewidth]{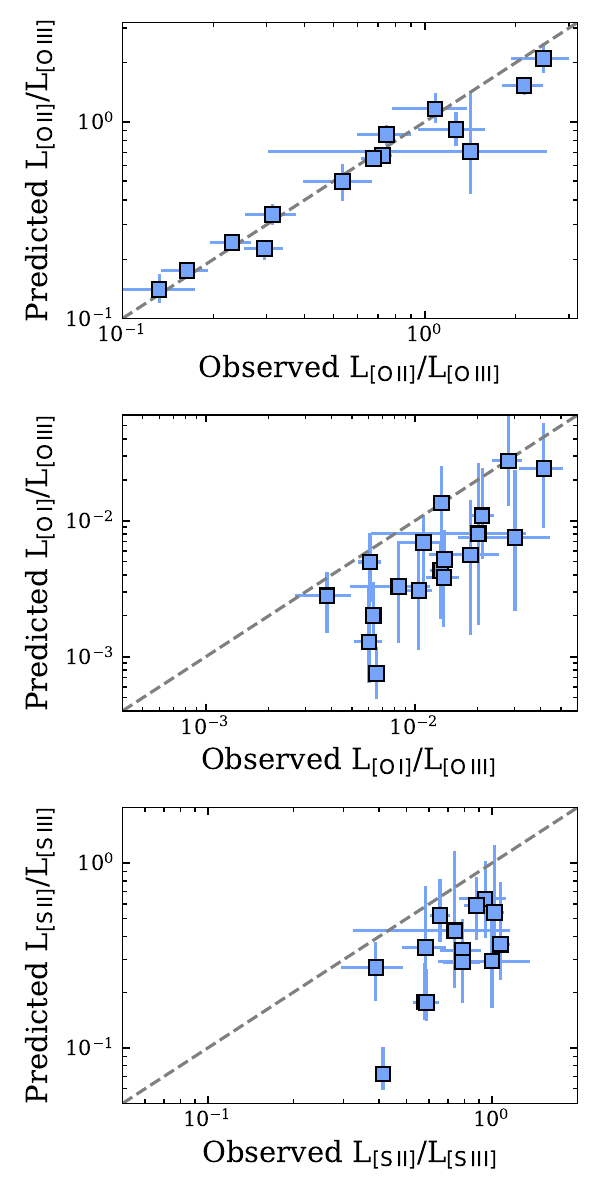}
    \caption{Comparisons of the \textsc{Cue} photoionization model emulator predictions vs. the observed values of the luminosity ratios of \oii$\lambda\lambda$3727,29/\oiii$\lambda\lambda$4960,5008 (top panel), \oi$\lambda$6302/\oiii$\lambda\lambda$4960,5008 (middle panel), and \sii$\lambda\lambda$6718,33/\siii$\lambda\lambda$9071,9533 (bottom panel). Dashed grey lines represent 1:1 lines. \textsc{Cue} accurately reproduces the luminosity of lines from ions with $E_{\textrm{ion}}>13.6$ eV, while it under-predicts the luminosity from \oi$\,$ and \sii, corroborating the perspective that some of the observed low-ionization emission may originate outside of \hii regions.}
    \label{fig:s23 and o23}
\end{figure}

\begin{figure}
    \centering
    \includegraphics[width=\linewidth]{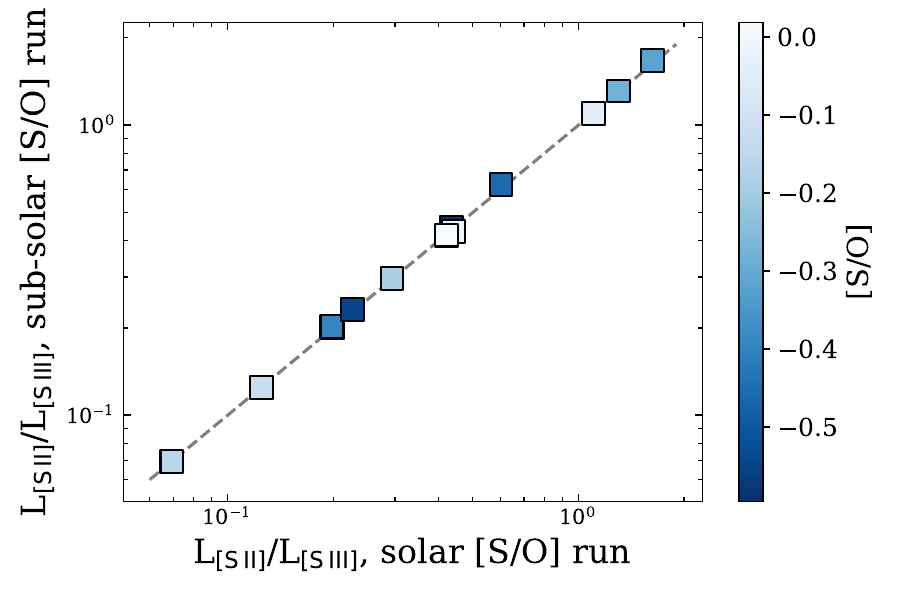}
    \caption{$L_{[\textrm{SII}]}/L_{[\textrm{SIII}]}$ predicted by \textsc{Cloudy} assuming solar abundances (x-axis) and altering log(S/H)  according to [S/O] from \protect{\cite{rogers2026}} (y-axis). The luminosity ratio depends primarily on ionization parameter and is largely invariant to changes in the sulfur abundance, thus we can use the \textsc{Cue} predictions of $L_{[\textrm{SII}]}/L_{[\textrm{SIII}]}$ to constrain the excess emission from \sii.}
    \label{fig:cloudycomp}
\end{figure}

In Figure \ref{fig:s23 and o23}, we compare observed CECILIA line luminosities with the representative model luminosities from \textsc{Cue}. We show line ratios to circumvent \textsc{Cue}'s assumption of solar S/O, which is inaccurate for CECILIA galaxies. While we expect \textsc{Cue} would overpredict \textit{individual} sulfur ion line luminosities, the ratio of two sulfur lines should be largely independent of S/O (we further consider this assumption below and in Figure \ref{fig:cloudycomp}).

In the top panel, we show the ratio of the line luminosities of singly-ionized oxygen O$^{+}$ ($E_{\textrm{ion}}=13.61$ eV) to doubly-ionized oxygen O$^{++}$ ($E_{\textrm{ion}}=35.11$ eV): $L_{\textrm{[OII]}\lambda\lambda3727,29}/L_{\textrm{[OIII]}\lambda\lambda4960,5008}$. These points scatter around the 1:1 (grey) line, showing that \textsc{Cue} accurately reproduces the relative emission of \oii$\,$ and \oiii. This is expected, as the ionization potentials of both ions are greater than 13.6 eV and therefore their emission should primarily originate in \hii regions. To illustrate the under-prediction of low-ionization lines in \textsc{Cue}, in the middle panel we show the ratio of the luminosities of neutral oxygen to doubly-ionized oxygen. \textsc{Cue} routinely under-predicts the luminosity of \oi$\lambda$6302, as may be expected if the majority of \oi$\,$ emission originates outside of an \hii region. 

 Akin to the under-prediction of \oi, in the bottom panel we show the ratio of the luminosities of singly-ionized sulfur S$^+$ ($E_{\textrm{ion}}=10.36$ eV) to doubly-ionized sulfur S$^{++}$ ($E_{\textrm{ion}}=23.33$ eV): $L_{\textrm{[SII]}\lambda\lambda6718,33 }/L_{\textrm{[SIII]}\lambda\lambda9071,9533}$. Here, the points also all lie below the 1:1 line. As the ionization potential of S$^{++}$ is between those of O$^+$ and O$^{++}$, both of which accurately predicted by \textsc{Cue}, \siii$\,$ emission should similarly be accurately predicted by the photoionization model. The under-prediction of $L_{[\textrm{S II}]}/L_{[\textrm{S III}]}$ is therefore assumed to be a consequence of the under-prediction of $L_{[\textrm{S II}]}$. 

The modeled ratio of $L_{[\textrm{S II}]}/L_{[\textrm{S III}]}$ is mainly dependent on log($U$), which is constrained in \textsc{Cue} with $L_{[\textrm{O II}]}/L_{[\textrm{O III}]}$. $L_{[\textrm{S II}]}/L_{[\textrm{S III}]}$ should not depend strongly on S/O \citep{kewley2019}, but we verify this by performing auxiliary \textsc{Cloudy} runs with sub-solar S/O. For each galaxy, we take  the maximum likelihood estimate (MLE) of the \textsc{Cue} posterior (ionizing spectrum, $U$, $n_{\textrm{H}}$, O/H, N/O) as the input to \textsc{Cloudy} and assume solar abundances from \cite{dopita2000} for the rest of the elements. \footnote{\textsc{Cue} was trained on \textsc{Cloudy} runs assuming \cite{dopita2000} abundances, thus for self-consistency we adopt them in these supplementary \textsc{Cloudy} runs.}
We use the "solar" run as a benchmark, and then alter the sulfur and argon abundances according to the values found for each galaxy in \cite{rogers2026} for a "sub-solar" run, accounting for any differences in the solar scales from \cite{asplund2009} and \cite{dopita2000}.
We compare the predicted $L_{\textrm{[SII]}}/L_{\textrm{[SIII]}}$ for both runs and find that $L_{\textrm{[SII]}}/L_{\textrm{[SIII]}}$ is consistent regardless of the S/O abundance  (Figure \ref{fig:cloudycomp}). This test confirms that the assumed S/O ratio does not impact the modeled $L_{\textrm{[SII]}}/L_{\textrm{[SIII]}}$ at fixed log($U$), $n_H$, and O/H; thus, we can use the difference between the observed and predicted $L_{\textrm{[SII]}}/L_{\textrm{[SIII]}}$ to constrain the \sii$\,$ luminosity originating outside of the star-forming regions.

To quantify the emission originating outside of \hii regions, we calculate a percent "excess emission": 

\begin{equation}
\label{eq-o1}
    f_{\textrm{excess,[OI]}} =100\times\frac{L_{\textrm{obs,[OI]}} - L_{\textrm{Cue, [OI]}}}{L_{\textrm{obs,[OI]}}}
\end{equation}
\begin{equation}
\label{eq-s2}
    f_{\textrm{excess,[SII]}} =100\times\frac{\frac{L_{\textrm{obs,[SII]}}}{L_{\textrm{obs,[SIII]}}} - \frac{L_{\textrm{Cue,[SII]}}}{L_{\textrm{Cue,[SIII]}}}}{\frac{L_{\textrm{obs,[SII]}}}{L_{\textrm{obs,[SIII]}}}}
\end{equation}
excess in the sense that there is some amount of observed emission \textit{in excess of} the photoionization model prediction. We estimate uncertainties on $f_{\textrm{excess}}$ using a Monte-Carlo sampling routine, sampling 1000 values taken from the normal distribution of the observed line luminosity and the \textsc{Cue}-predicted line luminosity distribution. We do not report values of excess emission in \sii$\,$ for galaxies that do not have predicted $L_{\textrm{[OII]}}/L_{\textrm{[OIII]}}$ within 1$\sigma$ of the observed value (RK120, BX523), or have non-detections ($< 3\sigma$) in any of the eight lines in $L_{\textrm{[OII]}}/L_{\textrm{[OIII]}}$ and $L_{\textrm{[SII]}}/L_{\textrm{[SIII]}}$  (BX216, fBM47, BX628, C31, BX274,  BX611, BX341, MD43).

\begin{figure*}
    \centering
    \includegraphics[width=1\textwidth]{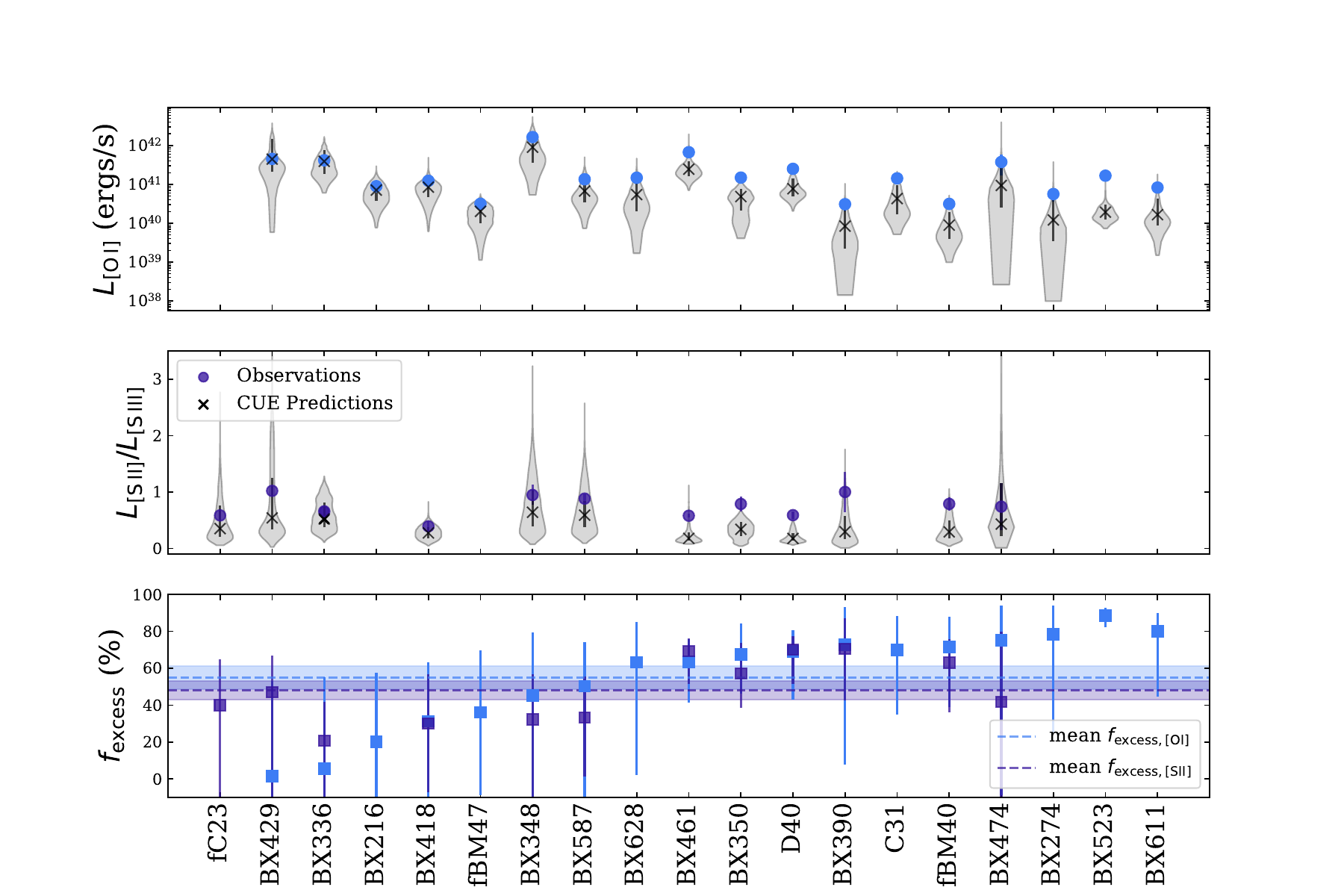}
    \caption{Comparison of observed luminosity vs. \textsc{Cue} luminosity distributions for \oi$\,$ (top panel) and \sii$\,$ (middle panel). The median of the luminosity distributions from \textsc{Cue} are marked by black Xs, with the 16th and 84th percentiles shown as errorbars. The bottom panel shows the percent of the observed luminosity ($f_{\textrm{excess}}$, eqs. \ref{eq-o1} and  \ref{eq-s2}) that is not reproduced by the photoionization model emulator for \oi$\,$ in blue and \sii$\,$ in dark purple. The mean excess emission for each species is shown as horizontal lines: $48.0\pm 5.1 \%$ for \sii$\,$ and $55.0 \pm 6.1 \%$ for \oi. Thus, about half of the low-ionization and neutral emission in CECILIA galaxies cannot be described by a typical \hii region model.}
    \label{fig:fout}
\end{figure*}

We calculate the excess fraction in \oi$\,$ for 18 galaxies, and the excess fraction in \sii$\,$ for 12 galaxies, as shown in Figure \ref{fig:fout}. In the top and middle panel we show the \textsc{Cue} predictions as grey violin distributions (median marked as black Xs) with the observed values as blue ($L_{\textrm{[OI]}}$) and dark purple ($L_{\textrm{[SII]}}/L_{\textrm{[SIII]}}$) points for comparison. The bottom panel shows the calculated \fx$\,$ for each line for each galaxy. The mean excess emission and associated standard error across the CECILIA sample in \sii$\,$ is $48.0\pm 5.1 \%$ and $55.0 \pm 6.1 \%$ in \oi$\,$ (dashed lines, bottom panel). These averages are formally consistent with one another, which may indicate that the excess low-ionization emission originates from the same emitting gas.

\begin{deluxetable*}{l|cc|cc|cc|cc|cc|cc}
\tablecaption{Spearman coefficients and $p$-values for $f_{\textrm{excess}}$ and galaxy properties \label{t:correlation}}
\tablewidth{\textwidth}
\tablehead{
  \colhead{} &
  \multicolumn{2}{c}{$L_{\textrm{H}\alpha}$} &
  \multicolumn{2}{c}{S2} &
  \multicolumn{2}{c}{O1} &
  \multicolumn{2}{c}{$n_e[\textrm{SII}]$} &
  \multicolumn{2}{c}{EW(H$\alpha$)} &
  \multicolumn{2}{c}{12+log(O/H)} \\
  \tableline
  & \colhead{$r$} & \multicolumn{1}{c|}{$p$}
    & \colhead{$r$} & \multicolumn{1}{c|}{$p$}
    & \colhead{$r$} & \multicolumn{1}{c|}{$p$}
    & \colhead{$r$} & \multicolumn{1}{c|}{$p$}
    & \colhead{$r$} & \multicolumn{1}{c|}{$p$}
    & \colhead{$r$} & \multicolumn{1}{c}{$p$}\\[-16pt]
}
\startdata
    $f_{\textrm{excess,[OI]}}$ & -0.36 & 0.14 & -0.11 & 0.67 & -0.01 & 0.96 & 0.11 & 0.66 & 0.21 & 0.40 & -0.41 & 0.11 \\
    \tableline
    $f_{\textrm{excess,[SII]}}$ & -0.36 & 0.26 & -0.02 & 0.95 & 0.20 & 0.56 & -0.12 & 0.71 & 0.17 & 0.59 & -0.46 & 0.15 \\
\enddata

\end{deluxetable*}

We next investigate the existence of correlations between $f_{\textrm{excess}}$ and galaxy properties in the CECILIA sample. In Table \ref{t:correlation} we show Spearman coefficients and $p$-values for $f_{\textrm{excess}}$ and galaxy properties including H$\alpha$ luminosity, S2, O1, $n_e$\sii, Ew(H$\alpha$), and 12+log(O/H). There are largely no significant correlations between $f_{\textrm{excess}}$ and H$\alpha$ luminosity, S2, O1, $n_e$\sii, EW(H$\alpha$) or gas-phase metallicity.

\section{Potential Sources of excess low-ionization emission in CECILIA galaxies}
\label{sec:emission}


\begin{figure*}
    \centering
    \includegraphics[width=\textwidth]{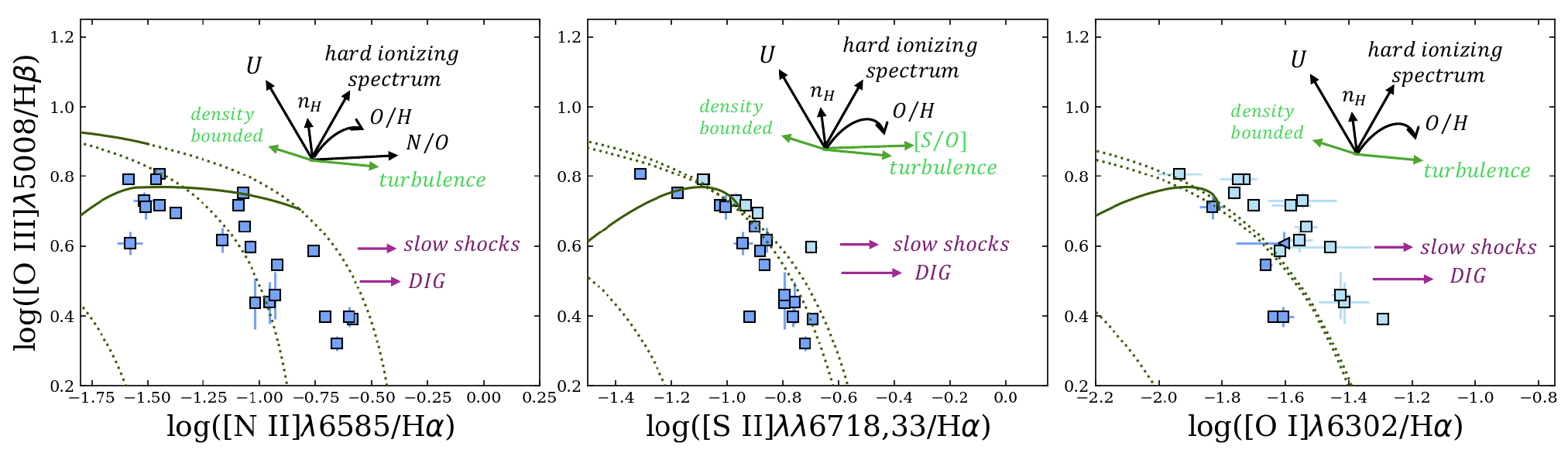}
    \caption{Representation of  factors capable of driving a mismatch between photoionization model predictions and galaxy-integrated spectra. Black arrows indicate ISM conditions that are variable in \textsc{Cue}, and green arrows indicate conditions that are not considered in \textsc{Cue}. Purple arrows indicate additional emission sources that will produce emission outside of standard \hii regions. CECILIA galaxies are given as blue data points with fiducial \textsc{Cloudy} models as dark green manifolds. Arrows are not intended to represent precise offsets, but rather approximations of the shape and relative magnitudes of the effects.}
    \label{fig:vector-bpt}
\end{figure*}

 The mismatch between the \cue models and the CECILIA galaxies further substantiates the perspective that there is some source of excess low-ionization emission in these galaxies not reproduced by a standard photoionization model. Investigating the source of this mismatch is complex; it is significantly computationally expensive to simulate all potential sources of emission in a galaxy. However, there are existing models and empirical observations of emission sources including shocks \citep{3mdb}, turbulence \citep{gray2017} and diffuse ionized gas \citep{zhang2017,shapley2019}, which can be compared with our galaxy observations to constrain their presence. However, the mismatch could also be from an unsuitable assumption in standard photoionization models like \textsc{Cue}. Different photoionization prescriptions (i.e., range of \textit{U} or ionizing spectrum) or gas abundance, geometry, and dynamics (i.e., range of O/H, $n_e$, choice of cloud bounding, turbulence) will alter the model's emission, and there may be some assumption made in \cue that is incorrect for Cosmic Noon galaxies. 

We illustrate these factors in Figure \ref{fig:vector-bpt}, offering a qualitative sense of how the location of a model or galaxy on the N2-, S2- and O1-BPTs can be moved. In this section, we first explore potential limitations of \textsc{Cue} when modeling Cosmic Noon galaxies, then attempt to constrain the presence of other sources of emission such as shocks, turbulence, and diffuse ionized gas in CECILIA by comparing our results to existing empirical relationships and models.



\subsection{Limitations of \textsc{Cue}}
\label{sec:interior}

As a photoionization model, \cue brings the benefit of not assuming particular underlying correlations between parameters; for instance, it does not prescribe a specific N/O-O/H relation, instead fitting the two properties independently. Further, it makes no assumption about the ionizing source, instead adopting a piecewise power-law informed by a broad range of modeled ionizing spectra. This is in contrast to most other photoionization codes which often must assign a certain stellar model, although it may vary in stellar metallicity or age \citep{perez-montero2014, valeasari2016, strom2018,2024Marconi}. The predicted emission lines are therefore unaffected by the choice of a stellar population or star-formation history, and the ionizing spectrum is informed only by the luminosities of the mid- and high-ionization lines that we assume originate in \hii regions.


As most low-ionization and neutral emission is produced at the edge of an \hii region, we first check that the stopping criteria does not truncate the cloud prematurely. The \textsc{Cue} photoionization model emulator was originally trained on \textsc{Cloudy} runs with a spherical geometry assuming an ionization-bounded \hii region with a constant density across the cloud as defined in \cite{byler2017}. The runs were stopped when the temperature reached 100 K or the electron density reached 10$\%$ of the hydrogen density. 
To assess the impact of these choices, we alter the stopping criteria in our representative \textsc{Cloudy} runs (used in Figure \ref{fig:cloudycomp}) to integrate until the electron density reaches 1$\%$ of the hydrogen density and find that, at most, this increases the \oi$\,$ luminosity by $\sim2\%$, while \sii$\,$ is generally increased by less than $1\%$.


Though we model an ionization-bounded region, we consider whether a density-bounded model may be more applicable at higher redshift, as higher observed star-formation rate surface densities (\sigsfr) \citep{beckman2000, alexandroff2015} may lead to a relative over-abundance of photons. Density-bounded \hii regions, however, typically have more emission from high-ionization species compared to low-ionization species \citep[leftward green arrow, Figure \ref{fig:vector-bpt};][]{stasinska2015, kewley2019}, whereas the CECILIA galaxies have elevated flux levels of low-ionization species. Thus we take the ionization-bounded model as a valid assumption; this is further bolstered by galaxy location on Figure \ref{fig:shocks}, as density-bounded models lie towards lower  \oi$\lambda$6302/\oiii$\lambda$5008. 

It may be more applicable to use a model that assumes a constant pressure in the \hii region rather than a constant density \citep{lehnert2009, kewley2019notthereview} to allow for a complex temperature and density structure. However, the original study using the photoionization models that \textsc{Cue} is based on \citep{byler2017} did not find a significant difference between models that assume a constant density vs. a constant pressure. We do not currently have sufficient coverage of density-sensitive ratios in multiple ionization zones to investigate density stratification within CECILIA galaxies, but we do note that 
the higher density predicted in \cue runs that include He I$\lambda$10833 may indicate that there is some density variation in these galaxies that is not fully captured by $n_e$\oii$\,$ and $n_e$\sii. Future observations of density-sensitive lines in other zones such as the high-ionization tracer C III]$\lambda\lambda$1907,09 \citep{topping2024, topping2025-electrondensity} will help further constrain this potential stratification in CECILIA galaxies. 

Altogether, we find that \textsc{Cue} is a suitable model for this work. Its advantage lies in that it allows for a flexible ionizing spectrum, and we find that the simplifying assumptions made are not a major driver of the mismatch between observed and predicted \oi$\,$ and \sii$\,$ emission. 


\subsection{Line Emission from Shocks and Turbulence at Cosmic Noon}
To avoid overly-expensive computation times, photoionization models generally assume a static ISM. This assumption, however, may not reflect conditions at high redshift: shock-producing supernova explosions and strong starburst-driven winds are prominent in galaxies with higher  star-formation rate surface densities \citep{ho2016}, such as those at high $z$. 
These phenomena can also lead to increased turbulence in star-forming clouds \citep{salim2015, westoby2026}, though the nature of interstellar turbulence at high redshift is not yet well-characterized. It is well-established that shocks can contribute to global emission-line spectra, 
 and that this emission will influence a galaxy's position on diagnostic diagrams in both the optical \citep[upper purple arrow, Figure \ref{fig:vector-bpt};][]{kewley2001, rich2010} and the UV \citep{mingozzi2024, flury2025}. Further, turbulence can alter the structure of a star-forming region, increasing its surface area and thereby decreasing the value of the ionization parameter at the expanding ionization front. A lower emergent ionization parameter increases the size of the transition layer between the fully-ionized gas and the surrounding neutral gas, boosting emission from \oii, \nii, \sii, and \oi$\,$ (rightward green arrow, Figure \ref{fig:vector-bpt}).


\cite{gray2017} examined how irradiating a turbulent ISM affects the location of a nebula on the N2-BPT, S2-BPT, and the O32-R23 diagrams. \footnote{$\mathrm{O32\equiv\log\left(\frac{[O\,II]\lambda\lambda3727,29} {[O\,III]\lambda\lambda4960,5008}\right)}$ \\$\mathrm{R23\equiv\log\left(\frac{[O\,II]\lambda\lambda3727,29 + [O\,III]\lambda\lambda4960,5008}{H\beta}\right)}$}
They found that supersonic turbulence driven on small physical scales most closely resembles the emission from galaxies at Cosmic Noon by comparing to the KBSS-LM1 stack from \cite{steidel2016} and to MOSDEF \citep[MOSFIRE Deep Evolution Field Survey;][]{sanders2016}. However, they found that the turbulent model closest to KBSS-LM1 in the N2-BPT and the O32-R23 diagram had a velocity $\sim$10-15 km/s lower than the model closest to the stack in the S2-BPT. Further, the closest model in the N2-BPT was $\sim0.6$ dex lower in S2 compared to the LM1 stack. While our CECILIA data points lie in the same region as the LM1 stack in the N2-BPT, they are $\sim0.15$ dex lower in the S2-BPT; However, this remains $\sim0.45$ dex away from the closest turbulent model in N2. Explaining the excess emission in S2 would therefore require higher-velocity turbulence, which is inconsistent with the position on other line-ratio diagrams. Further, these turbulent models were constructed with the assumption of a solar abundance pattern, and considering sub-solar S/O in the models would likely lead to a shift towards lower log(\sii/H$\alpha$) in the models, exacerbating this discrepancy. That the observed \nii $\,$ emission and O32 in CECILIA galaxies are accurately reproduced by \textsc{Cue} leads us to believe that turbulence within the star-forming region is not the main source of the excess emission in \sii$\,$ and \oi$\,$ we observe. 

As our understanding of turbulence in star-forming regions at high-redshift evolves with future high-resolution JWST spectroscopic studies, we will be able to further quantify how turbulence affects line emission from specific ionic species, and reveal if there are scenarios where turbulence preferentially boosts \sii$\,$ and \oi. Modeling the complex ISM is steadily developing: radiative transfer codes such as RAMSES-RTZ \citep{katz2022} 
are moving in this direction. In all models, it will be necessary to include the non-solar abundance patterns we see at high redshift.



\begin{figure}
    \centering
    \includegraphics[width=\linewidth]{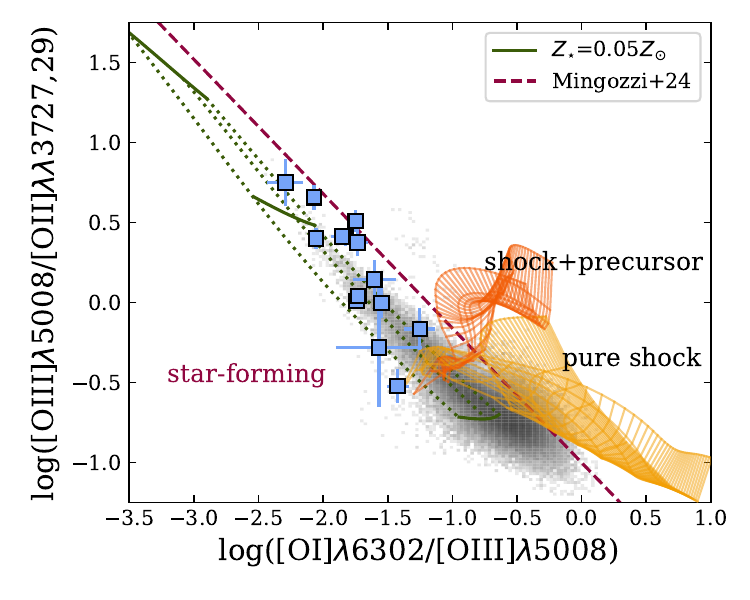}
    \caption{Location of CECILIA galaxies compared to \protect{\cite{allen2008}} shock models on the log(\oiii/\oii) vs. log(\oi/\oiii) diagram. The diagram is relatively insensitive to metallicity, allowing us to compare CECILIA galaxies with non-solar abundance patterns to shock models with LMC abundances. We see no significant overlap between CECILIA galaxies and the shock models, and CECILIA galaxies lie within the SF region defined by \protect{\cite{mingozzi2024}}, suggesting that shocks are not the main driver of the excess low-ionization emission.}
    \label{fig:shocks}
\end{figure}

 Quantifying the contribution from shocks in higher-redshift galaxies is similarly difficult, as current shock models do not yet account for the abundance patterns at high redshift. However, there are some parameter spaces that are relatively insensitive to metallicity, where we can analyze potential shock emission in CECILIA galaxies. In Figure \ref{fig:shocks}, we show CECILIA galaxies on the \oiii$\lambda$5008/\oii$\lambda\lambda$3727,29 vs \oi$\lambda$6302/\oiii$\lambda$5008 diagram, along with LMC-abundance (12+log(O/H) $ = 8.35$) shock models from \cite{allen2008}, made with the MAPPINGS code and accessed via the 3MdB database \citep{3mdb}. This diagram is relatively insensitive to gas-phase metallicity \citep{heckman1980, stasinska2015}, thus the different abundance patterns between the models and CECILIA galaxies should not significantly affect their comparison. We see that the CECILIA galaxies are located in the star-forming region as defined by \cite{mingozzi2024} and do not significantly overlap with the shock models including either a photoionization precursor (orange manifolds) or not (yellow manifolds).


 Further, CECILIA has recorded detections of several NIR lines, including the shock-sensitive line of \feii$\,$ \citep{olivia1989, calzetti1997, rosenberg2012}. Recent studies have proposed two "NIR-BPTs" incorporating this line to differentiate ionization sources in high-$z$ galaxies \citep[log(\heibpt)$\,$ or log(\siiibpt)$\,$ vs. log(\febpt);][]{brinchmann2023, calabro2023}. CECILIA galaxies occupy a similar locus as AURORA galaxies \citep[Figure 7 from][]{shapley2025} on these diagrams, with medians of log(\febpt)$\,$ $\sim-0.72$, log(\heibpt)$\,$$\sim0.49$, and log(\siiibpt)$\,$$\sim0.59$. This locus overlaps with photoionization models, providing additional evidence that shocks are not the main driver of the low-ionization emission mismatch in CECILIA, though we note that the CECILIA data points are offset towards higher \feii$\,$ than predicted by the BPASS $Z_\star=0.001$ models plotted in \cite{shapley2025}. However, there remain significant systematic uncertainties in the iron atomic data and assumed dust depletion in photoionization models.

From our existing rest optical and NIR data, there is no clear indication that shocks in CECILIA galaxies are driving the excess emission in \sii$\,$ and \oi. Unambiguous detection of shocks requires spectroscopy capable of resolving broad components typical of shocked gas. UV data can also clarify the presence of shocks, as shocks should produce very strong UV CELs, and certain line ratios including [CIII]$\lambda$1909/HeII$\lambda$1640 are lower in shocked regions than star-forming regions. In the future, the presence of shocks in CECILIA galaxies will be investigated using FUV spectra of individual galaxies recently acquired. In the same manner as as necessary for photoionization and turbulent models, it will be necessary to compile shock models that consider the non-solar abundance patterns at high $z$ as larger samples of high-redshift galaxies are assembled.





\subsection{Line Emission from Diffuse Ionized Gas at Cosmic Noon}

 As CECILIA galaxies show no clear evidence for strong contributions from shocks or turbulence, it is likely that, instead, this excess neutral and low-ionization emission may originate in diffuse ionized gas outside of \hii regions. This type of gas has been extensively studied in local-Universe galaxies with surveys such as MaNGA \citep{bundy2015}, CALIFA \citep{sanchez2012}, and PHANGS-MUSE \citep{phangs-muse2022}, and it is well-established that it can contribute to galaxy-integrated emission-line spectra (lower purple line, Figure \ref{fig:vector-bpt}). 

 DIG, as it is studied in the local Universe, has not one specific observational definition, but rather an array of characteristics. It has lower densities than \hii regions \citep{madsen2006} and its emission tends to be more prominent in regions with lower \sigsfr$\,$ \citep{oey2007a, zhang2017}. 
Based on this relationship with \sigsfr, it is generally assumed that the contribution from diffuse ionized gas to \ha$\,$ is low at $z>2$ \citep{shapley2019}, as high-$z$ star formation rate surface densities are considered too high for partially-ionized regions to be established.

The analysis of the presence of DIG emission in CECILIA galaxies differs from the low-$z$ concept of DIG in that we calculate the fraction of emission directly from \oi$\,$ and \sii$\,$ CELs, in order to constrain the excess emission specifically in these lines. To understand the characteristics of this potential diffuse source in CECILIA galaxies, we investigate if \fx$\,$ follows the same relationships as seen of DIG in the low-redshift Universe.

\begin{figure}
   \centering
    \includegraphics[width=\linewidth]{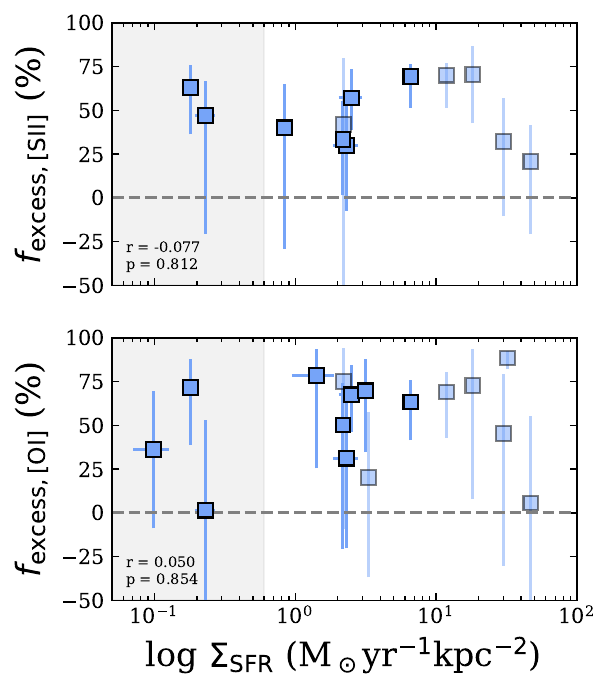}
    \caption{Relationship between excess emission in \sii$\,$ and \oi$\,$ and $\Sigma_{\textrm{SFR}}$ for CECILIA galaxies. Transparent points indicate galaxies with significant uncertainties ($\sigma_{R_e}>R_e$) on measured radii, uncertainties which we do not plot for visual clarity. Spearman coefficients considering all points are shown in the lower left. We see no significant trend between \sigsfr$\,$ and the amount of excess low-ionization emission in CECILIA galaxies, including when calculated with the subset of the most precisely-measured \sigsfr$\,$ values (opaque points),  indicating that CEL emission from diffuse ionized gas at high $z$ may not follow local empirical relationships. The grey shaded area highlights \sigsfr$\,$ values where this empirical low-$z$ relationship is defined with \ha.
    }
    \label{fig:diffuseionizedgas}
\end{figure}

In Figure \ref{fig:diffuseionizedgas} we show the relationship between \fx$\,$ and \sigsfr$\,$ for CECILIA galaxies. We calculate $\Sigma_{\textrm{SFR}}$ using SFR(H$\alpha$) determined using the metallicity-dependent conversion factors from \cite{korhonencuestas2025}. The half-light radii of the CECILIA galaxies are taken from existing imaging in F140W and F160W \citep{chen2021}. We see no significant trend between $f_{\textrm{excess}}$ and $\Sigma_{\textrm{SFR}}$ in the CECILIA sample, as confirmed by a Spearman's rank correlation test, indicating that galaxies with lower star-formation rate surface densities do not tend to have higher levels of excess \sii$\,$ or \oi$\,$ emission in our sample. This holds even when calculated only for galaxies with well-measured radii (i.e. $\sigma_{R_e}>R_e$; opaque points) or grouped by imaging filter. However, we note that this empirical relationship between \sigsfr$\,$ and DIG emission has mainly been established at $\sim\Sigma_{\textrm{SFR}}<0.6$ M$_\odot$ yr$^{-1}$ kpc$^{-2}$ (grey shaded region, Figure \ref{fig:diffuseionizedgas}), which the majority of CECILIA galaxies lie above. We also see no clear trend between the excess emission and other properties used to define DIG in the local Universe, including S2, O1, and EW(H$\alpha$) (Table \ref{t:correlation}).

 Though there is no clear relationship between \fx$\,$ and several observational definitions of low-$z$ DIG, it remains possible that the excess emission does originate in diffuse gas. As we reported in \cite{rogers2026}, the average $n_e$ in CECILIA as measured with \sii$\,$ is lower than that measured by \oii: $\langle n_e$\sii$\rangle=267\pm44 $ cm$^{-3}$ vs. $\langle n_e$\oii$\rangle=386\pm70$ cm$^{-3}$. If the excess \sii$\,$ emission does originate in lower-density gas, this may currently be biasing $n_e$[SII] low. Indeed, this measured density difference is seen in galaxy-integrated DESI spectra \citep{rong2026}.


With these $n_e$ averages and the average \fx$\,$ for \sii, we can roughly estimate the density of the gas producing the excess emission. To do this, we assume the \hii region density is described by only $n_e$[OII], and weight its contribution to $\frac{f_{6718}}{f_{6733}}$ by the amount we assume comes from \hii regions, then subtract this weighted \hii region contribution from the observed $\frac{f_{6718}}{f_{6733}}$ to get the weighted contribution from the "DIG," which is then used to calculate $n_e$:
\begin{flalign}
&\begin{aligned}
        \frac{f_{6718}}{f_{6733}}(n_e[\textrm{DIG}]) &= \frac{1}{\langle f_{\textrm{excess, [SII]}}\rangle}\times \\
    &\Biggl[\frac{f_{6718}}{f_{6733}}\bigl(\langle n_e[\textrm{SII}] \rangle \bigr) \times  \\
     &\biggl(\frac{f_{6718}}{f_{6733}}\bigl(\langle n_e[\textrm{OII}] \rangle\bigr)\times\bigl(1-\langle f_{\textrm{excess, [SII]}}\rangle\bigr)\biggr) \Biggr]
\end{aligned}&&
\end{flalign}
The ratio $\frac{f_{6718}}{f_{6733}}(n_e[\textrm{DIG}])$ is then input into \textsc{pyneb's} \texttt{getTemDen} function assuming $T_e=1.25\times10^4$K, and we recover a median $n_e$[DIG] of $\sim165^{+100}_{-79}$ cm$^{-3}$. This is generally much higher than local DIG electron densities, but this may not be inconsistent with the evolution of electron density seen at high redshift. This value is significantly lower than the \ion{H}{2}-representative $n_e$\oii, offering evidence for lower-density gas within CECILIA galaxies.

From these tests, we see that, though the \fx$\,$ does not follow empirical relationships that local DIG does, there may indeed be some lower-density ionized gas contributing to the global emission-line spectra of these galaxies. As galaxy structure is different at high-$z$, potentially lacking distinct boundaries between gas in star-forming regions and diffuse gas, DIG may evolve with redshift. Further investigation of the nature of DIG at high $z$ is necessary to truly constrain its presence, characteristics, and potential contribution to low-ionization CELs. The ideal testbed is spatially-resolved spectroscopy of distant galaxies capable of resolving sub-kpc scales. However, the typical spaxel size remains prohibitively large for this at high $z$. Considering current capabilities, lensed galaxies  \citep[such as the Sunburst arc;][]{riviera-thorson2026} are the best current opportunity to characterize high $z$ diffuse gas. Future investigation into the structure of the ISM at $z>1$ is also a promising science case for ELTs.


\section{Potential Bias in Sulfur Abundances}
\label{sec:effects}

As our analysis indicates that this excess emission in CECILIA galaxies is produced outside of \hii regions, we seek to constrain the potential bias introduced in abundance calculations when assuming all low-ionization emission is from \hii regions. 
Previous studies of MaNGA galaxies have shown that including emission from diffuse ionized gas in strong-line metallicity calculations does introduce a bias in metallicity measurements \citep{zhang2017, sanders2017, valeasari2019}. In \cite{rogers2026}  we did not consider emission from neutral oxygen to be from \hii regions. However, we did consider singly-ionized sulfur to originate within \hii regions.

To quantify how S/H and S/O may change if part of the \sii $\,$ emission is indeed produced in gas outside of \hii regions, we re-calculate S/H for each galaxy with a measured $f_{\textrm{excess, [SII]}}$. In this new calculation, only the luminosity of \sii$\lambda\lambda$6718,33 is altered; we scale its total luminosity by (1-$f_{\textrm{excess, [SII]}}$)
to isolate the luminosity we are confident originates in \hii regions. We then use this luminosity to re-calculate the S/H abundance and S/O abundance ratio using the same method as \cite{rogers2026}. In brief, we use \textsc{pyneb}'s \texttt{getIonAbundance} function with the relevant ionization zone $T_e$ and the averaged electron density from $n_e$\sii$\,$ and $n_e$\oii, or $n_e$\sii$\,$ if $n_e$\oii$\,$ is unavailable. The results are consistent when repeating the analysis using only $n_e$\oii, which should be more reflective of \hii regions.
To correct for the unobserved S$^{3+}$, we use Ionization Correction Factors (ICFs), derived from photoionization model predictions based on local star-forming galaxies from \cite{izotov2006}. As these ICFs are polynomial functions of O$^+$/(O$^+$+O$^{++}$), they will not be affected by the amount of \sii$\,$ assumed to be in \hii regions.

\begin{figure}
    \centering
    \includegraphics[width=\linewidth]{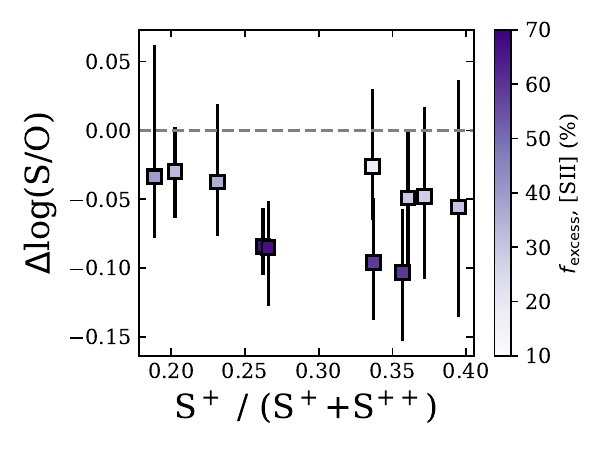}
    \caption{Offset from the original log(S/O) (y-axis) as a function of the relative contribution of S$^+$ to the total S/H calculation (x-axis). There is an overall trend towards an increasingly sub-solar log(S/O), enhanced when a greater fraction of S/H comes from S$^+$, or when a greater fraction of emission in S$^+$ is from outside the star-forming regions (darker purple points).}
    \label{fig:delta-so}
\end{figure}

The re-calculated abundances of S/H and S/O are, on average, $\sim0.06$ dex lower than the originally calculated abundances for this subset of galaxies.
While this is well within the uncertainty of the average log(S/O) in the CECILIA sample, in individual galaxies the difference between the original log(S/O) and the log(S/O) computed with emission only from \hii regions can be significant (Figure \ref{fig:delta-so}). 
 The offset from the original log(S/O) depends on how important S$^+$ is in the calculation of S/H, 
with the offset increasing as a larger fraction S$^+$/H$^+$ contributes to S/H. This offset is further exacerbated when a larger portion of the \sii$\,$ emission is not reproduced by the photoionization model (darker purple points, Figure \ref{fig:delta-so}). 

As more studies at Cosmic Noon and higher redshift calculate abundance ratios including sulfur, it is important to consider whether the galaxy-integrated emission of \sii$\,$ can be assumed to originate from the same \hii regions as \siii, as including emission from outside of star-forming regions will tend to bias S/O high. Luckily, Ar/O provides a complementary measurement that S/O can be compared to. Both argon and sulfur are thought to both be produced through Type Ia and CCSNe enrichment \citep{kobayashi2020,  foley2026}, 
and detectable emission lines of argon in the rest-optical are produced by ions with $E_{
\textrm{ion}}>13.6$ eV, therefore the measured Ar/O abundances should be uncontaminated by gas originating outside of \hii regions.

\section{Conclusions}
\label{sec:conclusions}

In this letter, we have analyzed neutral and low-ionization emission from \oi$\lambda$6302 and \sii$\lambda\lambda$6718,33 in a sample of 19 star-forming galaxies at $z\sim2-3$ observed as part of the CECILIA program, motivated by emerging knowledge of sub-solar sulfur abundance patterns at high-$z$ and new JWST detections of \oi$\lambda$6302. From this analysis, we draw several conclusions:
\begin{enumerate}
\item When analyzing strong emission lines on common diagnostic diagrams, it is necessary to account for non-solar abundance patterns (Figure \ref{fig:bpt-pi}). Not accounting for these physical conditions may obscure the presence of emission from additional sources beyond photoionization, specifically in \sii$\lambda\lambda$6718,33. 

\item To quantify the excess neutral and low-ionization emission in individual galaxies, we use the flexible photoionization model emulator \textsc{Cue} and find that a standard \hii region model that reproduces direct-method metallicities and mid- and high-ionization line emission cannot reproduce $48.0\pm 5.1 \%$ of \sii$\,$ emission and $55.0 \pm 6.1 \%$ of \oi$\,$ emission in the CECILIA sample (Figure \ref{fig:fout}). 

\item We find no clear evidence for turbulence or shocks (Figure \ref{fig:shocks}) driving this excess emission in CECILIA galaxies when comparing to existing models. However, these models are often calibrated against the low-$z$ Universe, and updated models that consider the distinct non-solar abundance patterns at high $z$ stand to offer further constraining power. 

\item Our findings imply that this excess emission may originate in diffuse ionized gas outside of \hii regions in these galaxies. However, we see that this gas does not behave similarly to DIG in the local Universe, following no established low-$z$ empirical relationships (Figure \ref{fig:diffuseionizedgas} and Table \ref{t:correlation}). Nevertheless, the difference between the average electron densities in CECILIA suggests this excess emission in \sii$\,$ does come from gas with a lower density than a typical high-$z$ \hii region, and we estimate a rough density of $n_e$[DIG] $\sim165^{+100}_{-79}$ cm$^{-3}$. Future spatially-resolved observations of low-ionization emission at higher redshift offer further constraining power on the nature of DIG at high $z$; the current most applicable testbed is lensed galaxies, though this is a promising science case for ELTs.

\item In the meantime, it is possible to correct for emission outside of star-forming regions in integrated galaxy spectra when calculating abundances. If there is no consideration that some \sii$\,$ emission may originate outside of \hii regions, sulfur abundances can be systematically biased high by $\sim0.06$ dex (Figure \ref{fig:delta-so}). 
For galaxies with a larger contribution from S$^{+}$ or more \sii$\,$ emission outside of \hii regions, the bias will be larger. However, Ar/O provides a complementary measure of the relative level of CCSN and Type Ia enrichment, which should be unaffected by emission originating outside of \hii regions. 
\end{enumerate}
\twocolumngrid

\begin{acknowledgements}
C.v.R would like to thank the 3MdB google group for access to the shock models used in this analysis, and Larrance Xing for his insights on turbulence at high-$z$. C.v.R also acknowledges the use of Github CoPilot (Microsoft) for aid in formatting figures; all scientific content, analysis, and conclusions are the authors. N. K. C. is supported by the National Science Foundation Graduate Research Fellowship Program under Grant No. 2025381248. G.C.R., R.F.T, and A.L.S. acknowledge partial support from the JWST-GO-02593.006-A, JWST-GO-02593.008-A, and JWST-GO-02593.004-A grants, respectively. N.S.J.R. was also supported by JWST-GO-02593.008-A. These funds were provided by NASA through a grant from the Space Telescope Science Institute, which is operated by the Association of Universities for Research in Astronomy, Inc., under NASA contract NAS503127. R.F.T. also acknowledges support from the Pittsburgh Foundation (grant ID UN2021-121482) and the Research Corporation for Scientific Advancement (Cottrell Scholar Award, grant ID 28289). A.L.S. is also supported by the David and Lucile Packard Foundation (Packard Fellowship, grant 2024-77399) and the National Science Foundation (grant number AST-2406780). GCR acknowledges the support from Grant 63667 from the John Templeton Foundation. Any opinions, findings, and conclusions or recommendations expressed in this material are those of the authors and do not necessarily reflect the views of the John Templeton Foundation. Y.L. is supported by the National Science Foundation (grant number AST-2406780) and a Research Corporation for Scientific Advancement (RCSA) Scialog Award (grant number SA-LSST-2024-094a). Z.Z. acknowledges the financial support from the CIERA Postdoctoral Fellowship.

\par 

This work is primarily based on observations made with NASA/ESA/CSA JWST, associated with PID 2593, which can be accessed via doi:\dataset[10.17909/x66z-p144]{https://doi.org/10.17909/x66z-p144}. The data were obtained from the Mikulski Archive for Space Telescopes (MAST) at the Space Telescope Science Institute, which is operated by the Association of Universities for Research in Astronomy, Inc., under NASA contract NAS 5-03127 for JWST. The ground-based spectroscopy included in the analysis were obtained at W.M. Keck Observatory, which is operated as a scientific partnership between the California Institute of Technology, the University of California, and NASA. Keck access was provided by NASA, the California Institute of Technology, as well as Northwestern University and the Center for Interdisciplinary Exploration and Research in Astrophysics (CIERA). The Observatory was made possible by the generous financial support of the W. M. Keck Foundation. We wish to thank Mauna a W$\bar{\textrm{a}}$kea for its gift of these data, and we seek to work toward an astronomical practice that honors the '$\bar{a}ina$ our telescopes reside on.

\end{acknowledgements}

\facilities{Keck:I (MOSFIRE), JWST (NIRSpec)}
\software{BPASSv2.2.1 \citep{stanway2018bpassv2.2.1}, Cloudy \citep{cloudy23}, \textsc{Cue} \cite{li2025-cue}, Matplotlib \citep{matplotlib}, NumPy \citep{numpy}, pandas \citep{pandas}, PyNeb \citep{luridiana2015}, SciPy \citep{scipy}}

\bibliography{ref_library}

@misc{korhonencuestas2026,
      title={CECILIA: Using Flexible Photoionization Models to Accurately Measure Abundances without Auroral Lines}, 
      author={Nathalie A. Korhonen Cuestas and Allison L. Strom and Yijia Li and Noah S. J. Rogers and Caroline von Raesfeld and Gwen C. Rudie and Ryan F. Trainor and Menelaos Raptis and Zhuyun Zhuang},
      year={2026},
      eprint={2609.22488},
      archivePrefix={arXiv},
      primaryClass={astro-ph.GA},
      url={https://arxiv.org/abs/2609.22488}, 
}

@software{scipy,
       author = {{Gommers}, Ralf and {Virtanen}, Pauli and {Haberland}, Matt and {Burovski}, Evgeni and {Reddy}, Tyler and {Weckesser}, Warren and {Oliphant}, Travis E. and {Nelson}, Andrew and {Cournapeau}, David and {alexbrc} and {Roy}, Pamphile and {Polat}, Ilhan and {Peterson}, Pearu and {Wilson}, Josh and {endolith} and {Mayorov}, Nikolay and {Colley}, Lucas and {van der Walt}, Stefan and {Brett}, Matthew and {Laxalde}, Denis and {Bowhay}, Jake and {Larson}, Eric and {Sakai}, Atsushi and {Millman}, Jarrod and {Lars} and {peterbell10} and {Carey}, CJ and {van Mulbregt}, Paul and {eric-jones} and {Steppi}, Albert},
        title = "{scipy/scipy: SciPy 1.15.3}",
         year = 2025,
        month = may,
          eid = {10.5281/zenodo.15366870},
          doi = {10.5281/zenodo.15366870},
      version = {v1.15.3},
    publisher = {Zenodo},
       adsurl = {https://ui.adsabs.harvard.edu/abs/2025zndo..15366870G}
}

@ARTICLE{matplotlib,
       author = {{Hunter}, John D.},
        title = "{Matplotlib: A 2D Graphics Environment}",
      journal = {Computing in Science and Engineering},
         year = 2007,
        month = jan,
       volume = {9},
       number = {3},
        pages = {90-95},
          doi = {10.1109/MCSE.2007.55},
       adsurl = {https://ui.adsabs.harvard.edu/abs/2007CSE.....9...90H}
}

@article{numpy,
	title = {Array programming with {NumPy}},
	volume = {585},
	copyright = {2020 The Author(s)},
	issn = {1476-4687},
	url = {https://www.nature.com/articles/s41586-020-2649-2},
	doi = {10.1038/s41586-020-2649-2},
	language = {en},
	number = {7825},
	urldate = {2026-09-18},
	journal = {Nature},
	publisher = {Nature Publishing Group},
	author = {Harris, Charles R. and Millman, K. Jarrod and van der Walt, Stéfan J. and Gommers, Ralf and Virtanen, Pauli and Cournapeau, David and Wieser, Eric and Taylor, Julian and Berg, Sebastian and Smith, Nathaniel J. and Kern, Robert and Picus, Matti and Hoyer, Stephan and van Kerkwijk, Marten H. and Brett, Matthew and Haldane, Allan and del Río, Jaime Fernández and Wiebe, Mark and Peterson, Pearu and Gérard-Marchant, Pierre and Sheppard, Kevin and Reddy, Tyler and Weckesser, Warren and Abbasi, Hameer and Gohlke, Christoph and Oliphant, Travis E.},
	month = sep,
	year = {2020},
	pages = {357--362},
}

@software{pandas,
  author       = {The pandas development team},
  title        = {pandas-dev/pandas: Pandas},
  month        = sep,
  year         = 2026,
  publisher    = {Zenodo},
  version      = {v3.0.6},
  doi          = {10.5281/zenodo.22820507},
  url          = {https://doi.org/10.5281/zenodo.22820507},
  swhid        = {swh:1:dir:fab735d044fd4c46ae48ed0c2a5c456286c476f5
                   ;origin=https://doi.org/10.5281/zenodo.3509134;vis
                   it=swh:1:snp:2696e0dbf27ceb94ff12bf90fd43c8178dd02
                   daf;anchor=swh:1:rel:bcdff04d4b5ba7a29d94777bbec88
                   ef823c37bbc;path=pandas-dev-pandas-619d761
                  },
}

@ARTICLE{cloudy22,
       author = {{Chatzikos}, M. and {Bianchi}, S. and {Camilloni}, F. and {Chakraborty}, P. and {Gunasekera}, C.~M. and {Guzm{\'a}n}, F. and {Milby}, J.~S. and {Sarkar}, A. and {Shaw}, G. and {van Hoof}, P.~A.~M. and {Ferland}, G.~J.},
        title = "{The 2023 Release of Cloudy}",
      journal = {\rmxaa},
         year = 2023,
        month = oct,
       volume = {59},
        pages = {327-343},
          doi = {10.22201/ia.01851101p.2023.59.02.12},
archivePrefix = {arXiv},
       eprint = {2308.06396},
 primaryClass = {astro-ph.GA},
       adsurl = {https://ui.adsabs.harvard.edu/abs/2023RMxAA..59..327C}
}

@ARTICLE{kewley2001agn,
       author = {{Kewley}, L.~J. and {Heisler}, C.~A. and {Dopita}, M.~A. and {Lumsden}, S.},
        title = "{Optical Classification of Southern Warm Infrared Galaxies}",
      journal = {\apjs},
         year = 2001,
        month = jan,
       volume = {132},
       number = {1},
        pages = {37-71},
          doi = {10.1086/318944},
       adsurl = {https://ui.adsabs.harvard.edu/abs/2001ApJS..132...37K}
}

@ARTICLE{rong2026,
       author = {{Rong}, Yu and {Liu}, Shihong and {Zou}, Hu},
        title = "{A DESI Calibration of the [O II]--[S II] Electron-density Offset in Integrated Star-forming Galaxies}",
      journal = {arXiv e-prints},
         year = 2026,
        month = jun,
          eid = {arXiv:2606.28129},
        pages = {arXiv:2606.28129},
          doi = {10.48550/arXiv.2606.28129},
archivePrefix = {arXiv},
       eprint = {2606.28129},
 primaryClass = {astro-ph.GA},
       adsurl = {https://ui.adsabs.harvard.edu/abs/2026arXiv260628129R}
}

@ARTICLE{roberts-borsani2024,
       author = {{Roberts-Borsani}, Guido and {Treu}, Tommaso and {Shapley}, Alice and {Fontana}, Adriano and {Pentericci}, Laura and {Castellano}, Marco and {Morishita}, Takahiro and {Bergamini}, Pietro and {Rosati}, Piero},
        title = "{Between the Extremes: A JWST Spectroscopic Benchmark for High-redshift Galaxies Using {\ensuremath{\sim}}500 Confirmed Sources at z {\ensuremath{\geq}} 5}",
      journal = {\apj},
         year = 2024,
        month = dec,
       volume = {976},
       number = {2},
          eid = {193},
        pages = {193},
          doi = {10.3847/1538-4357/ad85d3},
archivePrefix = {arXiv},
       eprint = {2403.07103},
 primaryClass = {astro-ph.GA},
       adsurl = {https://ui.adsabs.harvard.edu/abs/2024ApJ...976..193R}
}

@ARTICLE{topping2024,
       author = {{Topping}, Michael W. and {Stark}, Daniel P. and {Senchyna}, Peter and {Plat}, Adele and {Zitrin}, Adi and {Endsley}, Ryan and {Charlot}, St{\'e}phane and {Furtak}, Lukas J. and {Maseda}, Michael V. and {Smit}, Renske and {Mainali}, Ramesh and {Chevallard}, Jacopo and {Molyneux}, Stephen and {Rigby}, Jane R.},
        title = "{Metal-poor star formation at z > 6 with JWST: new insight into hard radiation fields and nitrogen enrichment on 20 pc scales}",
      journal = {\mnras},
         year = 2024,
        month = apr,
       volume = {529},
       number = {4},
        pages = {3301-3322},
          doi = {10.1093/mnras/stae682},
archivePrefix = {arXiv},
       eprint = {2401.08764},
 primaryClass = {astro-ph.GA},
       adsurl = {https://ui.adsabs.harvard.edu/abs/2024MNRAS.529.3301T}
}

@ARTICLE{scholte2026,
       author = {{Scholte}, D. and {Cullen}, F. and {Moustakas}, J. and {Zou}, H. and {Saintonge}, A. and {Arellano-Cordova}, K.~Z. and {Stanton}, T.~M. and {Andrews}, B. and {Sui}, J. and {Aguilar}, J. and {Ahlen}, S. and {Bianchi}, D. and {Brooks}, D. and {Castander}, F.~J. and {Cheng}, T. and {Claybaugh}, T. and {de la Macorra}, A. and {Dey}, B. and {Doel}, P. and {Douglass}, K. and {Ferraro}, S. and {Forero-Romero}, J.~E. and {Gazta{\~n}aga}, E. and {Gontcho}, S. Gontcho A. and {Gutierrez}, G. and {Joyce}, R. and {Kremin}, A. and {Lahav}, O. and {Landriau}, M. and {Le Guillou}, L. and {Martini}, P. and {Meisner}, A. and {Miquel}, R. and {Percival}, W.~J. and {Poppett}, C. and {Prada}, F. and {P{\'e}rez-R{\`a}fols}, I. and {Rossi}, G. and {Sanchez}, E. and {Schlegel}, D. and {Shao}, Z. and {Silber}, J. and {Sprayberry}, D. and {Tarl{\'e}}, G. and {Weaver}, B.~A.},
        title = "{Electron temperature relations and the direct N, O, Ne, S, and Ar abundances of 49 959 star-forming galaxies in DESI data release 2}",
      journal = {\mnras},
         year = 2026,
        month = sep,
       volume = {551},
       number = {1},
          eid = {stag1381},
        pages = {stag1381},
          doi = {10.1093/mnras/stag1381},
archivePrefix = {arXiv},
       eprint = {2601.02463},
 primaryClass = {astro-ph.GA},
       adsurl = {https://ui.adsabs.harvard.edu/abs/2026MNRAS.551g1381S}
}

@ARTICLE{stasinska2015,
       author = {{Stasi{\'n}ska}, G. and {Izotov}, Yu. and {Morisset}, C. and {Guseva}, N.},
        title = "{Excitation properties of galaxies with the highest [O iii]/[O ii] ratios. No evidence for massive escape of ionizing photons}",
      journal = {\aap},
         year = 2015,
        month = apr,
       volume = {576},
          eid = {A83},
        pages = {A83},
          doi = {10.1051/0004-6361/201425389},
archivePrefix = {arXiv},
       eprint = {1503.00320},
 primaryClass = {astro-ph.GA},
       adsurl = {https://ui.adsabs.harvard.edu/abs/2015A&A...576A..83S}
}

@ARTICLE{heckman1980,
       author = {{Heckman}, T.~M.},
        title = "{An Optical and Radio Survey of the Nuclei of Bright Galaxies - Activity in the Normal Galactic Nuclei}",
      journal = {\aap},
         year = 1980,
        month = jul,
       volume = {87},
        pages = {152},
       adsurl = {https://ui.adsabs.harvard.edu/abs/1980A&A....87..152H}
}

@ARTICLE{davies2021,
       author = {{Davies}, Rebecca L. and {F{\"o}rster Schreiber}, N.~M. and {Genzel}, R. and {Shimizu}, T.~T. and {Davies}, R.~I. and {Schruba}, A. and {Tacconi}, L.~J. and {{\"U}bler}, H. and {Wisnioski}, E. and {Wuyts}, S. and {Fossati}, M. and {Herrera-Camus}, R. and {Lutz}, D. and {Mendel}, J.~T. and {Naab}, T. and {Price}, S.~H. and {Renzini}, A. and {Wilman}, D. and {Beifiori}, A. and {Belli}, S. and {Burkert}, A. and {Chan}, J. and {Contursi}, A. and {Fabricius}, M. and {Lee}, M.~M. and {Saglia}, R.~P. and {Sternberg}, A.},
        title = "{The KMOS$^{3D}$ Survey: Investigating the Origin of the Elevated Electron Densities in Star-forming Galaxies at 1 {\ensuremath{\lesssim}} z {\ensuremath{\lesssim}} 3}",
      journal = {\apj},
         year = 2021,
        month = mar,
       volume = {909},
       number = {1},
          eid = {78},
        pages = {78},
          doi = {10.3847/1538-4357/abd551},
archivePrefix = {arXiv},
       eprint = {2012.10445},
 primaryClass = {astro-ph.GA},
       adsurl = {https://ui.adsabs.harvard.edu/abs/2021ApJ...909...78D}
}

@ARTICLE{kashino2017,
       author = {{Kashino}, D. and {Silverman}, J.~D. and {Sanders}, D. and {Kartaltepe}, J.~S. and {Daddi}, E. and {Renzini}, A. and {Valentino}, F. and {Rodighiero}, G. and {Juneau}, S. and {Kewley}, L.~J. and {Zahid}, H.~J. and {Arimoto}, N. and {Nagao}, T. and {Chu}, J. and {Sugiyama}, N. and {Civano}, F. and {Ilbert}, O. and {Kajisawa}, M. and {Le F{\`e}vre}, O. and {Maier}, C. and {Masters}, D. and {Miyaji}, T. and {Onodera}, M. and {Puglisi}, A. and {Taniguchi}, Y.},
        title = "{The FMOS-COSMOS Survey of Star-forming Galaxies at z {\ensuremath{\approx}} 1.6. IV. Excitation State and Chemical Enrichment of the Interstellar Medium}",
      journal = {\apj},
         year = 2017,
        month = jan,
       volume = {835},
       number = {1},
          eid = {88},
        pages = {88},
          doi = {10.3847/1538-4357/835/1/88},
archivePrefix = {arXiv},
       eprint = {1604.06802},
 primaryClass = {astro-ph.GA},
       adsurl = {https://ui.adsabs.harvard.edu/abs/2017ApJ...835...88K}
}

@ARTICLE{kaasinen2017,
       author = {{Kaasinen}, Melanie and {Bian}, Fuyan and {Groves}, Brent and {Kewley}, Lisa J. and {Gupta}, Anshu},
        title = "{The COSMOS-[O II] survey: evolution of electron density with star formation rate}",
      journal = {\mnras},
         year = 2017,
        month = mar,
       volume = {465},
       number = {3},
        pages = {3220-3234},
          doi = {10.1093/mnras/stw2827},
archivePrefix = {arXiv},
       eprint = {1611.01166},
 primaryClass = {astro-ph.GA},
       adsurl = {https://ui.adsabs.harvard.edu/abs/2017MNRAS.465.3220K}
}

@ARTICLE{3mdb,
       author = {{Alarie}, A. and {Morisset}, C.},
        title = "{Extensive Online Shock Model Database}",
      journal = {\rmxaa},
         year = 2019,
        month = oct,
       volume = {55},
        pages = {377-394},
          doi = {10.22201/ia.01851101p.2019.55.02.21},
archivePrefix = {arXiv},
       eprint = {1908.08579},
 primaryClass = {astro-ph.GA},
       adsurl = {https://ui.adsabs.harvard.edu/abs/2019RMxAA..55..377A}
}

@ARTICLE{rosenberg2012,
       author = {{Rosenberg}, M.~J.~F. and {van der Werf}, P.~P. and {Israel}, F.~P.},
        title = "{[FeII] as a tracer of supernova rate in nearby starburst galaxies}",
      journal = {\aap},
         year = 2012,
        month = apr,
       volume = {540},
          eid = {A116},
        pages = {A116},
          doi = {10.1051/0004-6361/201218772},
archivePrefix = {arXiv},
       eprint = {1202.2713},
 primaryClass = {astro-ph.CO},
       adsurl = {https://ui.adsabs.harvard.edu/abs/2012A&A...540A.116R}
}

@ARTICLE{calzetti1997,
       author = {{Calzetti}, Daniela},
        title = "{Reddening and Star Formation in Starburst Galaxies}",
      journal = {\aj},
         year = 1997,
        month = jan,
       volume = {113},
        pages = {162-184},
          doi = {10.1086/118242},
archivePrefix = {arXiv},
       eprint = {astro-ph/9610184},
 primaryClass = {astro-ph},
       adsurl = {https://ui.adsabs.harvard.edu/abs/1997AJ....113..162C}
}

@ARTICLE{olivia1989,
       author = {{Oliva}, E. and {Moorwood}, A.~F.~M. and {Danziger}, I.~J.},
        title = "{Infrared Spectroscopy of Supernova Remnants}",
      journal = {\aap},
         year = 1989,
        month = apr,
       volume = {214},
        pages = {307},
       adsurl = {https://ui.adsabs.harvard.edu/abs/1989A&A...214..307O}
}

@article{strom2022,
  title = {Chemical {{Abundance Scaling Relations}} for {{Multiple Elements}} in z {$\simeq$} 2--3 {{Star-forming Galaxies}}},
  author = {Strom, Allison L. and Rudie, Gwen C. and Steidel, Charles C. and Trainor, Ryan F.},
  year = 2022,
  month = feb,
  journal = {The Astrophysical Journal},
  volume = {925},
  number = {2},
  pages = {116},
  issn = {0004-637X, 1538-4357},
  doi = {10.3847/1538-4357/ac38a3},
  urldate = {2026-06-05},
  langid = {english}
}

@ARTICLE{flury2025,
       author = {{Flury}, Sophia R. and {Arellano-C{\'o}rdova}, Karla Z. and {Moran}, Edward C. and {Einsig}, Alaina},
        title = "{New ionization models and the shocking nitrogen excess at z > 5}",
      journal = {\mnras},
         year = 2025,
        month = nov,
       volume = {543},
       number = {4},
        pages = {3367-3381},
          doi = {10.1093/mnras/staf1615},
archivePrefix = {arXiv},
       eprint = {2412.06763},
 primaryClass = {astro-ph.GA},
       adsurl = {https://ui.adsabs.harvard.edu/abs/2025MNRAS.543.3367F}
}

@ARTICLE{mingozzi2024,
       author = {{Mingozzi}, Matilde and {James}, Bethan L. and {Berg}, Danielle A. and {Arellano-C{\'o}rdova}, Karla Z. and {Plat}, Adele and {Scarlata}, Claudia and {Aloisi}, Alessandra and {Amor{\'\i}n}, Ricardo O. and {Brinchmann}, Jarle and {Charlot}, St{\'e}phane and {Chisholm}, John and {Feltre}, Anna and {Gazagnes}, Simon and {Hayes}, Matthew and {Heckman}, Timothy and {Hernandez}, Svea and {Kewley}, Lisa J. and {Kumari}, Nimisha and {Leitherer}, Claus and {Martin}, Crystal L. and {Maseda}, Michael and {Nanayakkara}, Themiya and {Ravindranath}, Swara and {Rigby}, Jane R. and {Senchyna}, Peter and {Skillman}, Evan D. and {Sugahara}, Yuma and {Wilkins}, Stephen M. and {Wofford}, Aida and {Xu}, Xinfeng},
        title = "{CLASSY. VIII. Exploring the Source of Ionization with UV Interstellar Medium Diagnostics in Local High-z Analogs}",
      journal = {\apj},
         year = 2024,
        month = feb,
       volume = {962},
       number = {1},
          eid = {95},
        pages = {95},
          doi = {10.3847/1538-4357/ad1033},
archivePrefix = {arXiv},
       eprint = {2306.15062},
 primaryClass = {astro-ph.GA},
       adsurl = {https://ui.adsabs.harvard.edu/abs/2024ApJ...962...95M}
}

@ARTICLE{reddy2026,
       author = {{Reddy}, Naveen A. and {Shapley}, Alice E. and {Sanders}, Ryan L. and {Topping}, Michael W. and {Ellis}, Richard S. and {Pettini}, Max and {Brammer}, Gabriel and {Cullen}, Fergus and {F{\"o}rster Schreiber}, Natascha M. and {Khostovan}, Ali A. and {McLeod}, Derek J. and {McLure}, Ross J. and {Narayanan}, Desika and {Oesch}, Pascal A. and {Pahl}, Anthony J. and {Steidel}, Charles C. and {Berg}, Danielle A.},
        title = "{The AURORA Survey: Multiple Balmer and Paschen Emission Lines for Individual Star-forming Galaxies at z = 1.5─4.4. I. A Diversity of Nebular Attenuation Curves and Evidence for Non-unity Dust Covering Fractions}",
      journal = {\apj},
         year = 2026,
        month = mar,
       volume = {999},
       number = {1},
          eid = {15},
        pages = {15},
          doi = {10.3847/1538-4357/ae38da},
archivePrefix = {arXiv},
       eprint = {2506.17396},
 primaryClass = {astro-ph.GA},
       adsurl = {https://ui.adsabs.harvard.edu/abs/2026ApJ...999...15R}
}

@ARTICLE{sanders2025,
       author = {{Sanders}, Ryan L. and {Shapley}, Alice E. and {Topping}, Michael W. and {Reddy}, Naveen A. and {Berg}, Danielle A. and {Bouwens}, Rychard J. and {Brammer}, Gabriel and {Carnall}, Adam C. and {Cullen}, Fergus and {Dav{\'e}}, Romeel and {Dunlop}, James S. and {Ellis}, Richard S. and {F{\"o}rster Schreiber}, N.~M. and {Furlanetto}, Steven R. and {Glazebrook}, Karl and {Illingworth}, Garth D. and {Jones}, Tucker and {Kriek}, Mariska and {McLeod}, Derek J. and {McLure}, Ross J. and {Narayanan}, Desika and {Oesch}, Pascal A. and {Pahl}, Anthony J. and {Pettini}, Max and {Schaerer}, Daniel and {Stark}, Daniel P. and {Steidel}, Charles C. and {Tang}, Mengtao and {Clarke}, Leonardo and {Donnan}, Callum T. and {Kehoe}, Emily},
        title = "{The AURORA Survey: The Nebular Attenuation Curve of a Galaxy at z = 4.41 from Ultraviolet to Near-infrared Wavelengths}",
      journal = {\apj},
         year = 2025,
        month = aug,
       volume = {989},
       number = {2},
          eid = {209},
        pages = {209},
          doi = {10.3847/1538-4357/adf066},
archivePrefix = {arXiv},
       eprint = {2408.05273},
 primaryClass = {astro-ph.GA},
       adsurl = {https://ui.adsabs.harvard.edu/abs/2025ApJ...989..209S}
}

@ARTICLE{salim2015,
       author = {{Salim}, Diane M. and {Federrath}, Christoph and {Kewley}, Lisa J.},
        title = "{A Universal, Turbulence-regulated Star Formation Law: From Milky Way Clouds to High-redshift Disk and Starburst Galaxies}",
      journal = {\apjl},
         year = 2015,
        month = jun,
       volume = {806},
       number = {2},
          eid = {L36},
        pages = {L36},
          doi = {10.1088/2041-8205/806/2/L36},
archivePrefix = {arXiv},
       eprint = {1505.03144},
 primaryClass = {astro-ph.GA},
       adsurl = {https://ui.adsabs.harvard.edu/abs/2015ApJ...806L..36S}
}

@ARTICLE{westoby2026,
       author = {{Westoby}, B.~A. and {Hodge}, J.~A. and {Sharda}, P. and {Mancera Pi{\~n}a}, P.~E. and {Rybak}, M. and {da Cunha}, E. and {Li}, J. and {Smail}, I. and {Swinbank}, A.~M. and {Battisti}, A. and {Boogaard}, L.~A. and {Brandt}, W.~N. and {Calistro Rivera}, G. and {Chen}, C.-C. and {Cox}, P. and {Cracraft}, M. and {Dannerbauer}, H. and {Decarli}, R. and {Greve}, T.~R. and {Kendrew}, S. and {Knudsen}, K. and {Liao}, C.-L. and {van Marrewijk}, J. and {Nayak}, O. and {Neeleman}, M. and {Rowland}, L.~E. and {Schinnerer}, E. and {Walter}, F. and {Wardlow}, J.~L. and {Weiss}, A. and {van der Werf}, P.},
        title = "{Investigating the role of turbulence in the interstellar medium in $z\sim3$ dusty star-forming galaxies using kpc-resolution ALMA dust and gas maps}",
      journal = {arXiv e-prints},
         year = 2026,
        month = jun,
          eid = {arXiv:2606.11444},
        pages = {arXiv:2606.11444},
          doi = {10.48550/arXiv.2606.11444},
archivePrefix = {arXiv},
       eprint = {2606.11444},
 primaryClass = {astro-ph.GA},
       adsurl = {https://ui.adsabs.harvard.edu/abs/2026arXiv260611444W}
}

@ARTICLE{schaerer2026,
       author = {{Schaerer}, D. and {Izotov}, Y.~I. and {Marques-Chaves}, R. and {Steidel}, C.~C. and {Reddy}, N. and {Shapley}, A.~E. and {Mascia}, S. and {Chisholm}, J. and {Flury}, S.~R. and {Guseva}, N. and {Heckman}, T. and {Henry}, A. and {Inoue}, A.~K. and {Jung}, I. and {Kusakabe}, H. and {Mawatari}, K. and {Oesch}, P. and {{\"O}stlin}, G. and {Pentericci}, L. and {Roy}, N. and {Saldana-Lopez}, A. and {Sato}, R. and {Vanzella}, E. and {Verhamme}, A. and {Wang}, B.},
        title = "{Nitrogen abundances in star-forming galaxies 2.2 Gyr after the Big Bang are not elevated}",
      journal = {\aap},
         year = 2026,
        month = apr,
       volume = {708},
          eid = {A242},
        pages = {A242},
          doi = {10.1051/0004-6361/202556832},
archivePrefix = {arXiv},
       eprint = {2601.06968},
 primaryClass = {astro-ph.GA},
       adsurl = {https://ui.adsabs.harvard.edu/abs/2026A&A...708A.242S}
}

@ARTICLE{martinez2025,
       author = {{Martinez}, Zorayda and {Berg}, Danielle A. and {James}, Bethan L. and {Arellano-C{\'o}rdova}, Karla Z. and {Stark}, Daniel P. and {Senchyna}, Peter and {Skillman}, Evan D. and {Rogers}, Noah S.~J. and {Chisholm}, John},
        title = "{Under Pressure: Decoding the Effect of High Densities on Derived Nebular Properties}",
      journal = {\apj},
         year = 2025,
        month = dec,
       volume = {995},
       number = {2},
          eid = {204},
        pages = {204},
          doi = {10.3847/1538-4357/ae17c6},
archivePrefix = {arXiv},
       eprint = {2510.21960},
 primaryClass = {astro-ph.GA},
       adsurl = {https://ui.adsabs.harvard.edu/abs/2025ApJ...995..204M}
}

@ARTICLE{rogers2026-yp,
       author = {{Rogers}, Noah S.~J. and {Skillman}, Evan D. and {Pogge}, Richard W. and {Aver}, Erik and {Weller}, Miqaela K. and {Berg}, Danielle A. and {Salzer}, John J. and {Miller}, Jr, John H. and {Speigel}, Jayde and {Strom}, Allison L.},
        title = "{The LBT $Y_{\rm p}$ Project II: MODS Spectra, Physical Conditions, and Oxygen Abundances in Local Metal-Poor Nebulae}",
      journal = {arXiv e-prints},
         year = 2026,
        month = jan,
          eid = {arXiv:2601.22236},
        pages = {arXiv:2601.22236},
          doi = {10.48550/arXiv.2601.22236},
archivePrefix = {arXiv},
       eprint = {2601.22236},
 primaryClass = {astro-ph.GA},
       adsurl = {https://ui.adsabs.harvard.edu/abs/2026arXiv260122236R}
}

@ARTICLE{skillman2026,
       author = {{Skillman}, Evan D. and {Pogge}, Richard W. and {Aver}, Erik and {Rogers}, Noah S.~J. and {Weller}, Miqaela K. and {Olive}, Keith A. and {Berg}, Danielle A. and {Salzer}, John J. and {Miller}, Jr, John H. and {Spiegel}, Jayde and {Yeh}, Tsung-Han and {Fields}, Brian D.},
        title = "{The LBT $Y_{\rm p}$ Project I: An Improved Determination of the Primordial Helium Abundance -- Project Description, Sample Selection, Observations, and Methodology}",
      journal = {arXiv e-prints},
         year = 2026,
        month = jan,
          eid = {arXiv:2601.22232},
        pages = {arXiv:2601.22232},
          doi = {10.48550/arXiv.2601.22232},
archivePrefix = {arXiv},
       eprint = {2601.22232},
 primaryClass = {astro-ph.CO},
       adsurl = {https://ui.adsabs.harvard.edu/abs/2026arXiv260122232S}
}

@ARTICLE{berg2026,
       author = {{Berg}, Danielle A. and {Sanders}, Ryan L. and {Shapley}, Alice E. and {Topping}, Michael W. and {Reddy}, Naveen A. and {Skillman}, Evan D. and {Aver}, Erik and {Cullen}, Fergus and {Donnan}, Callum T. and {Dunlop}, James S. and {Jones}, Tucker and {Khostovan}, Ali Ahmad and {McLeod}, Derek J. and {Narayanan}, Desika and {Oesch}, Pascal A. and {Pahl}, Anthony J. and {Pettini}, Max and {Schreiber}, N.~M. F{\"o}rster and {Stark}, Daniel P.},
        title = "{The AURORA Survey: Robust Helium Abundances at High Redshift Reveal a Subpopulation of Helium-enhanced Galaxies in the Early Universe}",
      journal = {\apj},
         year = 2026,
        month = jan,
       volume = {996},
       number = {1},
          eid = {68},
        pages = {68},
          doi = {10.3847/1538-4357/ae18db},
archivePrefix = {arXiv},
       eprint = {2507.17057},
 primaryClass = {astro-ph.GA},
       adsurl = {https://ui.adsabs.harvard.edu/abs/2026ApJ...996...68B}
}

@ARTICLE{izotov2014,
       author = {{Izotov}, Y.~I. and {Thuan}, T.~X. and {Guseva}, N.~G.},
        title = "{A new determination of the primordial He abundance using the He I {\ensuremath{\lambda}}10830 {\r{A}} emission line: cosmological implications}",
      journal = {\mnras},
         year = 2014,
        month = nov,
       volume = {445},
       number = {1},
        pages = {778-793},
          doi = {10.1093/mnras/stu1771},
archivePrefix = {arXiv},
       eprint = {1408.6953},
 primaryClass = {astro-ph.CO},
       adsurl = {https://ui.adsabs.harvard.edu/abs/2014MNRAS.445..778I}
}

@ARTICLE{aver2015,
       author = {{Aver}, Erik and {Olive}, Keith A. and {Skillman}, Evan D.},
        title = "{The effects of He I {\ensuremath{\lambda}}10830 on helium abundance determinations}",
      journal = {\jcap},
         year = 2015,
        month = jul,
       volume = {2015},
       number = {7},
        pages = {011-011},
          doi = {10.1088/1475-7516/2015/07/011},
archivePrefix = {arXiv},
       eprint = {1503.08146},
 primaryClass = {astro-ph.CO},
       adsurl = {https://ui.adsabs.harvard.edu/abs/2015JCAP...07..011A}
}

@ARTICLE{katz2022,
       author = {{Katz}, Harley},
        title = "{RAMSES-RTZ: non-equilibrium metal chemistry and cooling coupled to on-the-fly radiation hydrodynamics}",
      journal = {\mnras},
         year = 2022,
        month = may,
       volume = {512},
       number = {1},
        pages = {348-365},
          doi = {10.1093/mnras/stac423},
archivePrefix = {arXiv},
       eprint = {2202.04083},
 primaryClass = {astro-ph.GA},
       adsurl = {https://ui.adsabs.harvard.edu/abs/2022MNRAS.512..348K}
}

@ARTICLE{rich2010,
       author = {{Rich}, J.~A. and {Dopita}, M.~A. and {Kewley}, L.~J. and {Rupke}, D.~S.~N.},
        title = "{NGC 839: Shocks in an M82-like Superwind}",
      journal = {\apj},
         year = 2010,
        month = sep,
       volume = {721},
       number = {1},
        pages = {505-517},
          doi = {10.1088/0004-637X/721/1/505},
archivePrefix = {arXiv},
       eprint = {1007.3495},
 primaryClass = {astro-ph.CO},
       adsurl = {https://ui.adsabs.harvard.edu/abs/2010ApJ...721..505R}
}

@ARTICLE{valeasari2019,
       author = {{Vale Asari}, N. and {Couto}, G.~S. and {Cid Fernandes}, R. and {Stasi{\'n}ska}, G. and {de Amorim}, A.~L. and {Ruschel-Dutra}, D. and {Werle}, A. and {Florido}, T.~Z.},
        title = "{Diffuse ionized gas and its effects on nebular metallicity estimates of star-forming galaxies}",
      journal = {\mnras},
         year = 2019,
        month = nov,
       volume = {489},
       number = {4},
        pages = {4721-4733},
          doi = {10.1093/mnras/stz2470},
archivePrefix = {arXiv},
       eprint = {1907.08635},
 primaryClass = {astro-ph.GA},
       adsurl = {https://ui.adsabs.harvard.edu/abs/2019MNRAS.489.4721V}
}

@ARTICLE{ho2016,
       author = {{Ho}, I.-Ting and {Medling}, Anne M. and {Bland-Hawthorn}, Joss and {Groves}, Brent and {Kewley}, Lisa J. and {Kobayashi}, Chiaki and {Dopita}, Michael A. and {Leslie}, Sarah K. and {Sharp}, Rob and {Allen}, James T. and {Bourne}, Nathan and {Bryant}, Julia J. and {Cortese}, Luca and {Croom}, Scott M. and {Dunne}, Loretta and {Fogarty}, L.~M.~R. and {Goodwin}, Michael and {Green}, Andy W. and {Konstantopoulos}, Iraklis S. and {Lawrence}, Jon S. and {Lorente}, Nuria P.~F. and {Owers}, Matt S. and {Richards}, Samuel and {Sweet}, Sarah M. and {Tescari}, Edoardo and {Valiante}, Elisabetta},
        title = "{The SAMI Galaxy Survey: extraplanar gas, galactic winds and their association with star formation history}",
      journal = {\mnras},
         year = 2016,
        month = apr,
       volume = {457},
       number = {2},
        pages = {1257-1278},
          doi = {10.1093/mnras/stw017},
archivePrefix = {arXiv},
       eprint = {1601.02022},
 primaryClass = {astro-ph.GA},
       adsurl = {https://ui.adsabs.harvard.edu/abs/2016MNRAS.457.1257H}
}

@ARTICLE{riviera-thorson2026,
       author = {{Rivera-Thorsen}, T. Emil and {Welch}, Brian and {Hutchison}, Taylor and {Hayes}, Matthew J. and {Rigby}, Jane R. and {Kim}, Keunho and {Choe}, Suhyeon and {Florian}, Michael and {Bayliss}, Matthew B. and {Khullar}, Gourav and {Sharon}, Keren and {Dahle}, H{\r{a}}kon and {Chisholm}, John and {Solhaug}, Erik and {Owens}, M. Riley and {Gladders}, Michael D.},
        title = "{The Sunburst Arc with JWST. IV. The Importance of Interaction, Turbulence, and Feedback for Lyman-continuum Escape}",
      journal = {\apj},
         year = 2026,
        month = apr,
       volume = {1000},
       number = {2},
          eid = {204},
        pages = {204},
          doi = {10.3847/1538-4357/ae4c3a},
archivePrefix = {arXiv},
       eprint = {2510.11702},
 primaryClass = {astro-ph.GA},
       adsurl = {https://ui.adsabs.harvard.edu/abs/2026ApJ..1000..204R}
}

@ARTICLE{alexandroff2015,
       author = {{Alexandroff}, Rachael M. and {Heckman}, Timothy M. and {Borthakur}, Sanchayeeta and {Overzier}, Roderik and {Leitherer}, Claus},
        title = "{Indirect Evidence for Escaping Ionizing Photons in Local Lyman Break Galaxy Analogs}",
      journal = {\apj},
         year = 2015,
        month = sep,
       volume = {810},
       number = {2},
          eid = {104},
        pages = {104},
          doi = {10.1088/0004-637X/810/2/104},
archivePrefix = {arXiv},
       eprint = {1504.02446},
 primaryClass = {astro-ph.GA},
       adsurl = {https://ui.adsabs.harvard.edu/abs/2015ApJ...810..104A}
}

@ARTICLE{beckman2000,
       author = {{Beckman}, J.~E. and {Rozas}, M. and {Zurita}, A. and {Watson}, R.~A. and {Knapen}, J.~H.},
        title = "{Populations of High-Luminosity Density-bounded H II Regions in Spiral Galaxies: Evidence and Implications}",
      journal = {\aj},
         year = 2000,
        month = jun,
       volume = {119},
       number = {6},
        pages = {2728-2744},
          doi = {10.1086/301380},
archivePrefix = {arXiv},
       eprint = {astro-ph/0003359},
 primaryClass = {astro-ph},
       adsurl = {https://ui.adsabs.harvard.edu/abs/2000AJ....119.2728B}
}

@misc{cleri2026,
      title={RUBIES: The Evolution of the Ionization Parameter from 0 < z < 9}, 
      author={Nikko J. Cleri and Zach J. Lewis and Joel Leja and Jakob M. Helton and Emilie Burnham and Olivia Curtis and Anna de Graaff and Michaela Hirschmann and Harley Katz and Michael V. Maseda and Ian McConachie and Adele Plat and Lucie Scharre},
      year={2026},
      eprint={2605.30410},
      archivePrefix={arXiv},
      primaryClass={astro-ph.GA},
      url={https://arxiv.org/abs/2605.30410}, 
}

@ARTICLE{hayes2025,
       author = {{Hayes}, Matthew J. and {Saldana-Lopez}, Alberto and {Citro}, Annalisa and {James}, Bethan L. and {Mingozzi}, Matilde and {Scarlata}, Claudia and {Martinez}, Zorayda and {Berg}, Danielle A.},
        title = "{On the Average Ultraviolet Emission-line Spectra of High-redshift Galaxies: Hot and Cold, Carbon-poor, Nitrogen Modest, and Oozing Ionizing Photons}",
      journal = {\apj},
         year = 2025,
        month = mar,
       volume = {982},
       number = {1},
          eid = {14},
        pages = {14},
          doi = {10.3847/1538-4357/adaea1},
archivePrefix = {arXiv},
       eprint = {2411.09262},
 primaryClass = {astro-ph.GA},
       adsurl = {https://ui.adsabs.harvard.edu/abs/2025ApJ...982...14H}
}

@ARTICLE{tang2025-ionization,
       author = {{Tang}, Mengtao and {Stark}, Daniel P. and {Plat}, Ad{\`e}le and {Feltre}, Anna and {Katz}, Harley and {Senchyna}, Peter and {Mason}, Charlotte A. and {Whitler}, Lily and {Chen}, Zuyi and {Topping}, Michael W.},
        title = "{JWST/NIRSpec Observations of High-ionization Emission Lines in Galaxies at High Redshift}",
      journal = {\apj},
         year = 2025,
        month = oct,
       volume = {991},
       number = {2},
          eid = {217},
        pages = {217},
          doi = {10.3847/1538-4357/adfd57},
archivePrefix = {arXiv},
       eprint = {2505.06359},
 primaryClass = {astro-ph.GA},
       adsurl = {https://ui.adsabs.harvard.edu/abs/2025ApJ...991..217T}
}

@ARTICLE{mascia2023,
       author = {{Mascia}, S. and {Pentericci}, L. and {Calabr{\`o}}, A. and {Treu}, T. and {Santini}, P. and {Yang}, L. and {Napolitano}, L. and {Roberts-Borsani}, G. and {Bergamini}, P. and {Grillo}, C. and {Rosati}, P. and {Vulcani}, B. and {Castellano}, M. and {Boyett}, K. and {Fontana}, A. and {Glazebrook}, K. and {Henry}, A. and {Mason}, C. and {Merlin}, E. and {Morishita}, T. and {Nanayakkara}, T. and {Paris}, D. and {Roy}, N. and {Williams}, H. and {Wang}, X. and {Brammer}, G. and {Brada{\v{c}}}, M. and {Chen}, W. and {Kelly}, P.~L. and {Koekemoer}, A.~M. and {Trenti}, M. and {Windhorst}, R.~A.},
        title = "{Closing in on the sources of cosmic reionization: First results from the GLASS-JWST program}",
      journal = {\aap},
         year = 2023,
        month = apr,
       volume = {672},
          eid = {A155},
        pages = {A155},
          doi = {10.1051/0004-6361/202345866},
archivePrefix = {arXiv},
       eprint = {2301.02816},
 primaryClass = {astro-ph.GA},
       adsurl = {https://ui.adsabs.harvard.edu/abs/2023A&A...672A.155M}
}

@ARTICLE{topping2025-ionization,
       author = {{Topping}, Michael W. and {Stark}, Daniel P. and {Senchyna}, Peter and {Chen}, Zuyi and {Zitrin}, Adi and {Endsley}, Ryan and {Charlot}, St{\'e}phane and {Furtak}, Lukas J. and {Maseda}, Michael V. and {Plat}, Adele and {Smit}, Renske and {Mainali}, Ramesh and {Chevallard}, Jacopo and {Molyneux}, Stephen and {Rigby}, Jane R.},
        title = "{Deep Rest-UV JWST/NIRSpec Spectroscopy of Early Galaxies: The Demographics of C IV and N-emitters in the Reionization Era}",
      journal = {\apj},
         year = 2025,
        month = feb,
       volume = {980},
       number = {2},
          eid = {225},
        pages = {225},
          doi = {10.3847/1538-4357/ada95c},
archivePrefix = {arXiv},
       eprint = {2407.19009},
 primaryClass = {astro-ph.GA},
       adsurl = {https://ui.adsabs.harvard.edu/abs/2025ApJ...980..225T}
}

@ARTICLE{li2025-electrondensity,
       author = {{Li}, Sijia and {Yu}, Si-Yue and {Ho}, Luis C. and {Silverman}, John D. and {Wang}, Jing and {Saintonge}, Am{\'e}lie and {Yu}, Niankun and {Fei}, Qinyue and {Kashino}, Daichi and {Yu}, Hao-ran},
        title = "{Linking Electron Density with Elevated Star Formation Activity from z = 0 to z = 10}",
      journal = {\apjl},
         year = 2025,
        month = nov,
       volume = {993},
       number = {2},
          eid = {L51},
        pages = {L51},
          doi = {10.3847/2041-8213/ae1695},
archivePrefix = {arXiv},
       eprint = {2510.18764},
 primaryClass = {astro-ph.GA},
       adsurl = {https://ui.adsabs.harvard.edu/abs/2025ApJ...993L..51L}
}

@ARTICLE{topping2025-electrondensity,
       author = {{Topping}, Michael W. and {Sanders}, Ryan L. and {Shapley}, Alice E. and {Pahl}, Anthony J. and {Reddy}, Naveen A. and {Stark}, Daniel P. and {Berg}, Danielle A. and {Clarke}, Leonardo and {Cullen}, Fergus and {Dunlop}, James S. and {Ellis}, Richard S. and {Schreiber}, N.~M. F{\"o}rster and {Illingworth}, Garth D. and {Jones}, Tucker and {Narayanan}, Desika and {Pettini}, Max and {Schaerer}, Daniel},
        title = "{The AURORA survey: the evolution of multiphase electron densities at high redshift}",
      journal = {\mnras},
         year = 2025,
        month = aug,
       volume = {541},
       number = {2},
        pages = {1707-1721},
          doi = {10.1093/mnras/staf903},
archivePrefix = {arXiv},
       eprint = {2502.08712},
 primaryClass = {astro-ph.GA},
       adsurl = {https://ui.adsabs.harvard.edu/abs/2025MNRAS.541.1707T}
}

@ARTICLE{stanton2026,
       author = {{Stanton}, T.~M. and {Cullen}, F. and {Carnall}, A.~C. and {Scholte}, D. and {Arellano-C{\'o}rdova}, K.~Z. and {Shapley}, A.~E. and {McLeod}, D.~J. and {Donnan}, C.~T. and {Begley}, R. and {Dav{\'e}}, R. and {Dunlop}, J.~S. and {McLure}, R.~J. and {Rowlands}, K. and {Bondestam}, C. and {Hamadouche}, M.~L. and {Leung}, H.-H. and {Stevenson}, S.~D. and {Taylor}, E.},
        title = "{The JWST EXCELS Survey: gas-phase metallicity evolution at 2 < z < 8}",
      journal = {\mnras},
         year = 2026,
        month = apr,
       volume = {547},
       number = {4},
          eid = {stag449},
        pages = {stag449},
          doi = {10.1093/mnras/stag449},
archivePrefix = {arXiv},
       eprint = {2511.00705},
 primaryClass = {astro-ph.GA},
       adsurl = {https://ui.adsabs.harvard.edu/abs/2026MNRAS.547ag449S}
}

@ARTICLE{tang2026,
       author = {{Tang}, Mengtao and {Stark}, Daniel P. and {Mason}, Charlotte A. and {Gelli}, Viola and {Chen}, Zuyi and {Topping}, Michael W.},
        title = "{The JWST Spectroscopic Properties of Galaxies at z = 9{\ensuremath{-}}14}",
      journal = {\apj},
         year = 2026,
        month = apr,
       volume = {1001},
       number = {1},
          eid = {38},
        pages = {38},
          doi = {10.3847/1538-4357/ae4edb},
archivePrefix = {arXiv},
       eprint = {2507.08245},
 primaryClass = {astro-ph.GA},
       adsurl = {https://ui.adsabs.harvard.edu/abs/2026ApJ..1001...38T}
}

@ARTICLE{pollock2026,
       author = {{Pollock}, Clara L. and {Gottumukkala}, Rashmi and {Heintz}, Kasper E. and {Brammer}, Gabriel B. and {Roberts-Borsani}, Guido and {Oesch}, Pascal A. and {Witstok}, Joris and {Arellano-C{\'o}rdova}, Karla Z. and {Cullen}, Fergus and {Scholte}, Dirk and {Terp}, Chamilla and {Rowland}, Lucie and {Sneppen}, Albert and {Ito}, Kei and {Valentino}, Francesco and {Matthee}, Jorryt and {Watson}, Darach and {Toft}, Sune},
        title = "{Novel z {\ensuremath{\sim}} 10 auroral line measurements extend the gradual offset of the fundamental metallicity relation deep into the first gigayear of cosmic time}",
      journal = {\aap},
         year = 2026,
        month = apr,
       volume = {708},
          eid = {A203},
        pages = {A203},
          doi = {10.1051/0004-6361/202556032},
archivePrefix = {arXiv},
       eprint = {2506.15779},
 primaryClass = {astro-ph.GA},
       adsurl = {https://ui.adsabs.harvard.edu/abs/2026A&A...708A.203P}
}

@ARTICLE{sarkar2025,
       author = {{Sarkar}, Arnab and {Chakraborty}, Priyanka and {Vogelsberger}, Mark and {McDonald}, Michael and {Torrey}, Paul and {Garcia}, Alex M. and {Khullar}, Gourav and {Ferland}, Gary J. and {Forman}, William and {Wolk}, Scott and {Schneider}, Benjamin and {Bautz}, Mark and {Miller}, Eric and {Grant}, Catherine and {ZuHone}, John},
        title = "{Unveiling the Cosmic Chemistry: Revisiting the Mass─Metallicity Relation with JWST/NIRSpec at 4 < z < 10}",
      journal = {\apj},
         year = 2025,
        month = jan,
       volume = {978},
       number = {2},
          eid = {136},
        pages = {136},
          doi = {10.3847/1538-4357/ad8f32},
archivePrefix = {arXiv},
       eprint = {2408.07974},
 primaryClass = {astro-ph.GA},
       adsurl = {https://ui.adsabs.harvard.edu/abs/2025ApJ...978..136S}
}

@ARTICLE{morishita2024,
       author = {{Morishita}, Takahiro and {Stiavelli}, Massimo and {Grillo}, Claudio and {Rosati}, Piero and {Schuldt}, Stefan and {Trenti}, Michele and {Bergamini}, Pietro and {Boyett}, Kit and {Chary}, Ranga-Ram and {Leethochawalit}, Nicha and {Roberts-Borsani}, Guido and {Treu}, Tommaso and {Vanzella}, Eros},
        title = "{Diverse Oxygen Abundance in Early Galaxies Unveiled by Auroral Line Analysis with JWST}",
      journal = {\apj},
         year = 2024,
        month = aug,
       volume = {971},
       number = {1},
          eid = {43},
        pages = {43},
          doi = {10.3847/1538-4357/ad5290},
archivePrefix = {arXiv},
       eprint = {2402.14084},
 primaryClass = {astro-ph.GA},
       adsurl = {https://ui.adsabs.harvard.edu/abs/2024ApJ...971...43M}
}

@ARTICLE{curti2024,
       author = {{Curti}, Mirko and {Maiolino}, Roberto and {Curtis-Lake}, Emma and {Chevallard}, Jacopo and {Carniani}, Stefano and {D'Eugenio}, Francesco and {Looser}, Tobias J. and {Scholtz}, Jan and {Charlot}, Stephane and {Cameron}, Alex and {{\"U}bler}, Hannah and {Witstok}, Joris and {Boyett}, Kristian and {Laseter}, Isaac and {Sandles}, Lester and {Arribas}, Santiago and {Bunker}, Andrew and {Giardino}, Giovanna and {Maseda}, Michael V. and {Rawle}, Tim and {Rodr{\'\i}guez Del Pino}, Bruno and {Smit}, Renske and {Willott}, Chris J. and {Eisenstein}, Daniel J. and {Hausen}, Ryan and {Johnson}, Benjamin and {Rieke}, Marcia and {Robertson}, Brant and {Tacchella}, Sandro and {Williams}, Christina C. and {Willmer}, Christopher and {Baker}, William M. and {Bhatawdekar}, Rachana and {Egami}, Eiichi and {Helton}, Jakob M. and {Ji}, Zhiyuan and {Kumari}, Nimisha and {Perna}, Michele and {Shivaei}, Irene and {Sun}, Fengwu},
        title = "{JADES: Insights into the low-mass end of the mass-metallicity-SFR relation at 3 < z < 10 from deep JWST/NIRSpec spectroscopy}",
      journal = {\aap},
         year = 2024,
        month = apr,
       volume = {684},
          eid = {A75},
        pages = {A75},
          doi = {10.1051/0004-6361/202346698},
archivePrefix = {arXiv},
       eprint = {2304.08516},
 primaryClass = {astro-ph.GA},
       adsurl = {https://ui.adsabs.harvard.edu/abs/2024A&A...684A..75C}
}

@ARTICLE{curti2023,
       author = {{Curti}, Mirko and {D'Eugenio}, Francesco and {Carniani}, Stefano and {Maiolino}, Roberto and {Sandles}, Lester and {Witstok}, Joris and {Baker}, William M. and {Bennett}, Jake S. and {Piotrowska}, Joanna M. and {Tacchella}, Sandro and {Charlot}, Stephane and {Nakajima}, Kimihiko and {Maheson}, Gabriel and {Mannucci}, Filippo and {Amiri}, Amirnezam and {Arribas}, Santiago and {Belfiore}, Francesco and {Bonaventura}, Nina R. and {Bunker}, Andrew J. and {Chevallard}, Jacopo and {Cresci}, Giovanni and {Curtis-Lake}, Emma and {Hayden-Pawson}, Connor and {Jones}, Gareth C. and {Kumari}, Nimisha and {Laseter}, Isaac and {Looser}, Tobias J. and {Marconi}, Alessandro and {Maseda}, Michael V. and {Scholtz}, Jan and {Smit}, Renske and {{\"U}bler}, Hannah and {Wallace}, Imaan E.~B.},
        title = "{The chemical enrichment in the early Universe as probed by JWST via direct metallicity measurements at z {\ensuremath{\sim}} 8}",
      journal = {\mnras},
         year = 2023,
        month = jan,
       volume = {518},
       number = {1},
        pages = {425-438},
          doi = {10.1093/mnras/stac2737},
archivePrefix = {arXiv},
       eprint = {2207.12375},
 primaryClass = {astro-ph.GA},
       adsurl = {https://ui.adsabs.harvard.edu/abs/2023MNRAS.518..425C}
}

@ARTICLE{nakajima2023,
       author = {{Nakajima}, Kimihiko and {Ouchi}, Masami and {Isobe}, Yuki and {Harikane}, Yuichi and {Zhang}, Yechi and {Ono}, Yoshiaki and {Umeda}, Hiroya and {Oguri}, Masamune},
        title = "{JWST Census for the Mass-Metallicity Star Formation Relations at z = 4-10 with Self-consistent Flux Calibration and Proper Metallicity Calibrators}",
      journal = {\apjs},
         year = 2023,
        month = dec,
       volume = {269},
       number = {2},
          eid = {33},
        pages = {33},
          doi = {10.3847/1538-4365/acd556},
archivePrefix = {arXiv},
       eprint = {2301.12825},
 primaryClass = {astro-ph.GA},
       adsurl = {https://ui.adsabs.harvard.edu/abs/2023ApJS..269...33N}
}

@ARTICLE{fujimoto2023,
       author = {{Fujimoto}, Seiji and {Arrabal Haro}, Pablo and {Dickinson}, Mark and {Finkelstein}, Steven L. and {Kartaltepe}, Jeyhan S. and {Larson}, Rebecca L. and {Burgarella}, Denis and {Bagley}, Micaela B. and {Behroozi}, Peter and {Chworowsky}, Katherine and {Hirschmann}, Michaela and {Trump}, Jonathan R. and {Wilkins}, Stephen M. and {Yung}, L.~Y. Aaron and {Koekemoer}, Anton M. and {Papovich}, Casey and {Pirzkal}, Nor and {Ferguson}, Henry C. and {Fontana}, Adriano and {Grogin}, Norman A. and {Grazian}, Andrea and {Kewley}, Lisa J. and {Kocevski}, Dale D. and {Lotz}, Jennifer M. and {Pentericci}, Laura and {Ravindranath}, Swara and {Somerville}, Rachel S. and {Wilkins}, Stephen M. and {Amor{\'\i}n}, Ricardo O. and {Backhaus}, Bren E. and {Calabr{\`o}}, Antonello and {Casey}, Caitlin M. and {Cooper}, M.~C. and {Fern{\'a}ndez}, Vital and {Franco}, Maximilien and {Giavalisco}, Mauro and {Hathi}, Nimish P. and {Harish}, Santosh and {Hutchison}, Taylor A. and {Iyer}, Kartheik G. and {Jung}, Intae and {Lucas}, Ray A. and {Zavala}, Jorge A.},
        title = "{CEERS Spectroscopic Confirmation of NIRCam-selected z {\ensuremath{\gtrsim}} 8 Galaxy Candidates with JWST/NIRSpec: Initial Characterization of Their Properties}",
      journal = {\apjl},
         year = 2023,
        month = jun,
       volume = {949},
       number = {2},
          eid = {L25},
        pages = {L25},
          doi = {10.3847/2041-8213/acd2d9},
archivePrefix = {arXiv},
       eprint = {2301.09482},
 primaryClass = {astro-ph.GA},
       adsurl = {https://ui.adsabs.harvard.edu/abs/2023ApJ...949L..25F}
}

@ARTICLE{heintz2023,
       author = {{Heintz}, Kasper E. and {Brammer}, Gabriel B. and {Gim{\'e}nez-Arteaga}, Clara and {Strait}, Victoria B. and {Lagos}, Claudia del P. and {Vijayan}, Aswin P. and {Matthee}, Jorryt and {Watson}, Darach and {Mason}, Charlotte A. and {Hutter}, Anne and {Toft}, Sune and {Fynbo}, Johan P.~U. and {Oesch}, Pascal A.},
        title = "{Dilution of chemical enrichment in galaxies 600 Myr after the Big Bang}",
      journal = {Nature Astronomy},
         year = 2023,
        month = dec,
       volume = {7},
        pages = {1517-1524},
          doi = {10.1038/s41550-023-02078-7},
archivePrefix = {arXiv},
       eprint = {2212.02890},
 primaryClass = {astro-ph.GA},
       adsurl = {https://ui.adsabs.harvard.edu/abs/2023NatAs...7.1517H}
}

@ARTICLE{abdurrouf2024,
       author = {{Abdurro'uf} and {Larson}, Rebecca L. and {Coe}, Dan and {Hsiao}, Tiger Yu-Yang and {{\'A}lvarez-M{\'a}rquez}, Javier and {G{\'o}mez}, Alejandro Crespo and {Adamo}, Angela and {Bhatawdekar}, Rachana and {Bik}, Arjan and {Bradley}, Larry D. and {Conselice}, Christopher J. and {Dayal}, Pratika and {Diego}, Jose M. and {Fujimoto}, Seiji and {Furtak}, Lukas J. and {Hutchison}, Taylor A. and {Jung}, Intae and {Killi}, Meghana and {Kokorev}, Vasily and {Mingozzi}, Matilde and {Norman}, Colin and {Resseguier}, Tom and {Ricotti}, Massimo and {Rigby}, Jane R. and {Vanzella}, Eros and {Welch}, Brian and {Windhorst}, Rogier A. and {Xu}, Xinfeng and {Zitrin}, Adi},
        title = "{JWST NIRSpec High-resolution Spectroscopy of MACS0647─JD at z = 10.167: Resolved [O II] Doublet and Electron Density in an Early Galaxy}",
      journal = {\apj},
         year = 2024,
        month = sep,
       volume = {973},
       number = {1},
          eid = {47},
        pages = {47},
          doi = {10.3847/1538-4357/ad6001},
archivePrefix = {arXiv},
       eprint = {2404.16201},
 primaryClass = {astro-ph.GA},
       adsurl = {https://ui.adsabs.harvard.edu/abs/2024ApJ...973...47A}
}

@ARTICLE{reddy2023,
       author = {{Reddy}, Naveen A. and {Topping}, Michael W. and {Sanders}, Ryan L. and {Shapley}, Alice E. and {Brammer}, Gabriel},
        title = "{A JWST/NIRSpec Exploration of the Connection between Ionization Parameter, Electron Density, and Star-formation-rate Surface Density in z = 2.7-6.3 Galaxies}",
      journal = {\apj},
         year = 2023,
        month = aug,
       volume = {952},
       number = {2},
          eid = {167},
        pages = {167},
          doi = {10.3847/1538-4357/acd754},
archivePrefix = {arXiv},
       eprint = {2303.11397},
 primaryClass = {astro-ph.GA},
       adsurl = {https://ui.adsabs.harvard.edu/abs/2023ApJ...952..167R}
}

@ARTICLE{isobe2023a,
       author = {{Isobe}, Yuki and {Ouchi}, Masami and {Nakajima}, Kimihiko and {Harikane}, Yuichi and {Ono}, Yoshiaki and {Xu}, Yi and {Zhang}, Yechi and {Umeda}, Hiroya},
        title = "{Redshift Evolution of Electron Density in the Interstellar Medium at z   0-9 Uncovered with JWST/NIRSpec Spectra and Line-spread Function Determinations}",
      journal = {\apj},
         year = 2023,
        month = oct,
       volume = {956},
       number = {2},
          eid = {139},
        pages = {139},
          doi = {10.3847/1538-4357/acf376},
archivePrefix = {arXiv},
       eprint = {2301.06811},
 primaryClass = {astro-ph.GA},
       adsurl = {https://ui.adsabs.harvard.edu/abs/2023ApJ...956..139I}
}

@ARTICLE{kobayashi2009,
       author = {{Kobayashi}, Chiaki and {Nomoto}, Ken'ichi},
        title = "{The Role of Type Ia Supernovae in Chemical Evolution. I. Lifetime of Type Ia Supernovae and Metallicity Effect}",
      journal = {\apj},
         year = 2009,
        month = dec,
       volume = {707},
       number = {2},
        pages = {1466-1484},
          doi = {10.1088/0004-637X/707/2/1466},
archivePrefix = {arXiv},
       eprint = {0801.0215},
 primaryClass = {astro-ph},
       adsurl = {https://ui.adsabs.harvard.edu/abs/2009ApJ...707.1466K}
}

@ARTICLE{clarke2026,
       author = {{Clarke}, Leonardo and {Lam}, Natalie and {Shapley}, Alice E. and {Topping}, Michael W. and {Brammer}, Gabriel B. and {Sanders}, Ryan L. and {Reddy}, Naveen A. and {Karthikeyan}, Shreya},
        title = "{Emission-line Diagnostics at z {\ensuremath{\gtrsim}} 2: A Probe of the Ionizing Spectrum and {\ensuremath{\alpha}}-enhancement beyond Cosmic Noon}",
      journal = {\apjl},
         year = 2026,
        month = may,
       volume = {1002},
       number = {1},
          eid = {L15},
        pages = {L15},
          doi = {10.3847/2041-8213/ae57a6},
       adsurl = {https://ui.adsabs.harvard.edu/abs/2026ApJ..1002L..15C}
}

@ARTICLE{isobe2026-jades,
       author = {{Isobe}, Yuki and {Maiolino}, Roberto and {Ji}, Xihan and {D'Eugenio}, Francesco and {Simmonds}, Charlotte and {Scholtz}, Jan and {Juod{\v{z}}balis}, Ignas and {Saxena}, Aayush and {Witstok}, Joris and {Kobayashi}, Chiaki and {Vanni}, Irene and {Salvadori}, Stefania and {Watanabe}, Kuria and {Monty}, Stephanie and {Belokurov}, Vasily and {Feltre}, Anna and {McClymont}, William and {Tacchella}, Sandro and {Curti}, Mirko and {{\"U}bler}, Hannah and {Charlot}, St{\'e}phane and {Bunker}, Andrew J. and {Chevallard}, Jacopo and {Curtis-Lake}, Emma and {Kumari}, Nimisha and {Rinaldi}, Pierluigi and {Robertson}, Brant and {Williams}, Christina C. and {Willott}, Chris},
        title = "{JADES: the chemical enrichment pattern of distant galaxies ─ {\ensuremath{\alpha}} enhancement, silicon depletion, and iron enhancement}",
      journal = {\mnras},
         year = 2026,
        month = apr,
       volume = {547},
       number = {3},
          eid = {stag123},
        pages = {stag123},
          doi = {10.1093/mnras/stag123},
archivePrefix = {arXiv},
       eprint = {2509.18055},
 primaryClass = {astro-ph.GA},
       adsurl = {https://ui.adsabs.harvard.edu/abs/2026MNRAS.547ag123I}
}

@ARTICLE{foley2026,
       author = {{Foley}, Jack and {Shapley}, Alice and {Sanders}, Ryan and {Reddy}, Naveen A. and {Topping}, Michael W. and {Stanton}, Thomas M. and {Pettini}, Max and {Cullen}, Fergus and {Ellis}, Richard S. and {Schreiber}, N.~M. F{\"o}rster and {Jones}, Tucker and {Pahl}, Anthony J. and {Clarke}, Leonardo and {Lam}, Natalie},
        title = "{The AURORA Survey: Constraining Chemical Enrichment Pathways at Cosmic Noon with Argon Abundances}",
      journal = {\apj},
         year = 2026,
        month = jul,
       volume = {1006},
       number = {1},
          eid = {54},
        pages = {54},
          doi = {10.3847/1538-4357/ae7c6d},
archivePrefix = {arXiv},
       eprint = {2512.10130},
 primaryClass = {astro-ph.GA},
       adsurl = {https://ui.adsabs.harvard.edu/abs/2026ApJ..1006...54F}
}

@ARTICLE{stanton2024-nirvandels,
       author = {{Stanton}, T.~M. and {Cullen}, F. and {McLure}, R.~J. and {Shapley}, A.~E. and {Arellano-C{\'o}rdova}, K.~Z. and {Begley}, R. and {Amor{\'\i}n}, R. and {Barrufet}, L. and {Calabr{\`o}}, A. and {Carnall}, A.~C. and {Cirasuolo}, M. and {Dunlop}, J.~S. and {Donnan}, C.~T. and {Hamadouche}, M.~L. and {Liu}, F.~Y. and {McLeod}, D.~J. and {Pentericci}, L. and {Pozzetti}, L. and {Sanders}, R.~L. and {Scholte}, D. and {Topping}, M.~W.},
        title = "{The NIRVANDELS survey: the stellar and gas-phase mass-metallicity relations of star-forming galaxies at z = 3.5}",
      journal = {\mnras},
         year = 2024,
        month = aug,
       volume = {532},
       number = {3},
        pages = {3102-3119},
          doi = {10.1093/mnras/stae1705},
archivePrefix = {arXiv},
       eprint = {2405.00774},
 primaryClass = {astro-ph.GA},
       adsurl = {https://ui.adsabs.harvard.edu/abs/2024MNRAS.532.3102S}
}

@ARTICLE{stanton2025-excels,
       author = {{Stanton}, T.~M. and {Cullen}, F. and {Carnall}, A.~C. and {Scholte}, D. and {Arellano-C{\'o}rdova}, K.~Z. and {McLeod}, D.~J. and {Begley}, R. and {Donnan}, C.~T. and {Dunlop}, J.~S. and {Hamadouche}, M.~L. and {McLure}, R.~J. and {Shapley}, A.~E. and {Bondestam}, C. and {Stevenson}, S.},
        title = "{The JWST EXCELS survey: tracing the chemical enrichment pathways of high-redshift star-forming galaxies with O, Ar, and Ne abundances}",
      journal = {\mnras},
         year = 2025,
        month = feb,
       volume = {537},
       number = {2},
        pages = {1735-1748},
          doi = {10.1093/mnras/staf106},
archivePrefix = {arXiv},
       eprint = {2411.11837},
 primaryClass = {astro-ph.GA},
       adsurl = {https://ui.adsabs.harvard.edu/abs/2025MNRAS.537.1735S}
}

@ARTICLE{runco2022,
       author = {{Runco}, Jordan N. and {Reddy}, Naveen A. and {Shapley}, Alice E. and {Steidel}, Charles C. and {Sanders}, Ryan L. and {Strom}, Allison L. and {Coil}, Alison L. and {Kriek}, Mariska and {Mobasher}, Bahram and {Pettini}, Max and {Rudie}, Gwen C. and {Siana}, Brian and {Topping}, Michael W. and {Trainor}, Ryan F. and {Freeman}, William R. and {Shivaei}, Irene and {Azadi}, Mojegan and {Price}, Sedona H. and {Leung}, Gene C.~K. and {Fetherolf}, Tara and {de Groot}, Laura and {Zick}, Tom and {Fornasini}, Francesca M. and {Barro}, Guillermo},
        title = "{Reconciling the results of the z   2 MOSDEF and KBSS-MOSFIRE Surveys}",
      journal = {\mnras},
         year = 2022,
        month = jul,
       volume = {513},
       number = {3},
        pages = {3871-3892},
          doi = {10.1093/mnras/stac1115},
archivePrefix = {arXiv},
       eprint = {2112.09715},
 primaryClass = {astro-ph.GA},
       adsurl = {https://ui.adsabs.harvard.edu/abs/2022MNRAS.513.3871R}
}

@ARTICLE{madsen2006,
       author = {{Madsen}, G.~J. and {Reynolds}, R.~J. and {Haffner}, L.~M.},
        title = "{A Multiwavelength Optical Emission Line Survey of Warm Ionized Gas in the Galaxy}",
      journal = {\apj},
         year = 2006,
        month = nov,
       volume = {652},
       number = {1},
        pages = {401-425},
          doi = {10.1086/508441},
archivePrefix = {arXiv},
       eprint = {astro-ph/0609558},
 primaryClass = {astro-ph},
       adsurl = {https://ui.adsabs.harvard.edu/abs/2006ApJ...652..401M}
}

@ARTICLE{phangs-muse2022,
       author = {{Emsellem}, Eric and {Schinnerer}, Eva and {Santoro}, Francesco and {Belfiore}, Francesco and {Pessa}, Ismael and {McElroy}, Rebecca and {Blanc}, Guillermo A. and {Congiu}, Enrico and {Groves}, Brent and {Ho}, I.-Ting and {Kreckel}, Kathryn and {Razza}, Alessandro and {Sanchez-Blazquez}, Patricia and {Egorov}, Oleg and {Faesi}, Chris and {Klessen}, Ralf S. and {Leroy}, Adam K. and {Meidt}, Sharon and {Querejeta}, Miguel and {Rosolowsky}, Erik and {Scheuermann}, Fabian and {Anand}, Gagandeep S. and {Barnes}, Ashley T. and {Be{\v{s}}li{\'c}}, Ivana and {Bigiel}, Frank and {Boquien}, M{\'e}d{\'e}ric and {Cao}, Yixian and {Chevance}, M{\'e}lanie and {Dale}, Daniel A. and {Eibensteiner}, Cosima and {Glover}, Simon C.~O. and {Grasha}, Kathryn and {Henshaw}, Jonathan D. and {Hughes}, Annie and {Koch}, Eric W. and {Kruijssen}, J.~M. Diederik and {Lee}, Janice and {Liu}, Daizhong and {Pan}, Hsi-An and {Pety}, J{\'e}r{\^o}me and {Saito}, Toshiki and {Sandstrom}, Karin M. and {Schruba}, Andreas and {Sun}, Jiayi and {Thilker}, David A. and {Usero}, Antonio and {Watkins}, Elizabeth J. and {Williams}, Thomas G.},
        title = "{The PHANGS-MUSE survey. Probing the chemo-dynamical evolution of disc galaxies}",
      journal = {\aap},
         year = 2022,
        month = mar,
       volume = {659},
          eid = {A191},
        pages = {A191},
          doi = {10.1051/0004-6361/202141727},
archivePrefix = {arXiv},
       eprint = {2110.03708},
 primaryClass = {astro-ph.GA},
       adsurl = {https://ui.adsabs.harvard.edu/abs/2022A&A...659A.191E}
}

@ARTICLE{2024Marconi,
       author = {{Marconi}, A. and {Amiri}, A. and {Feltre}, A. and {Belfiore}, F. and {Cresci}, G. and {Curti}, M. and {Mannucci}, F. and {Bertola}, E. and {Brazzini}, M. and {Carniani}, S. and {Cataldi}, E. and {D'Amato}, Q. and {de Rosa}, G. and {Di Teodoro}, E. and {Ginolfi}, M. and {Kumari}, N. and {Marconcini}, C. and {Maiolino}, R. and {Magrini}, L. and {Marasco}, A. and {Mingozzi}, M. and {Moreschini}, B. and {Nagao}, T. and {Oliva}, E. and {Scialpi}, M. and {Tomicic}, N. and {Tozzi}, G. and {Ulivi}, L. and {Venturi}, G.},
        title = "{HOMERUN: A new approach to photoionization modeling: I. Reproducing observed emission lines with percent accuracy and obtaining accurate physical properties of the ionized gas}",
      journal = {\aap},
         year = 2024,
        month = sep,
       volume = {689},
          eid = {A78},
        pages = {A78},
          doi = {10.1051/0004-6361/202449240},
archivePrefix = {arXiv},
       eprint = {2401.13028},
 primaryClass = {astro-ph.GA},
       adsurl = {https://ui.adsabs.harvard.edu/abs/2024A&A...689A..78M}
}

@INPROCEEDINGS{stasinska2025,
       author = {{Stasi{\'n}ska}, Gra{\.z}yna},
        title = "{On determining the chemical composition of planetary nebulae}",
    booktitle = {Planetary Nebulae: A Universal Toolbox in the Era of Precision Astrophysics},
         year = 2025,
       editor = {{De Marco}, O. and {Zijlstra}, A. and {Szczerba}, R.},
       series = {IAU Symposium},
       volume = {384},
        month = jan,
        pages = {122-135},
          doi = {10.1017/S1743921323005781},
archivePrefix = {arXiv},
       eprint = {2312.01873},
 primaryClass = {astro-ph.GA},
       adsurl = {https://ui.adsabs.harvard.edu/abs/2025IAUS..384..122S}
}

@ARTICLE{kobayashi2020,
       author = {{Kobayashi}, Chiaki and {Karakas}, Amanda I. and {Lugaro}, Maria},
        title = "{The Origin of Elements from Carbon to Uranium}",
      journal = {\apj},
         year = 2020,
        month = sep,
       volume = {900},
       number = {2},
          eid = {179},
        pages = {179},
          doi = {10.3847/1538-4357/abae65},
archivePrefix = {arXiv},
       eprint = {2008.04660},
 primaryClass = {astro-ph.GA},
       adsurl = {https://ui.adsabs.harvard.edu/abs/2020ApJ...900..179K}
}

@ARTICLE{chen2021,
       author = {{Chen}, Yuguang and {Steidel}, Charles C. and {Erb}, Dawn K. and {Law}, David R. and {Trainor}, Ryan F. and {Reddy}, Naveen A. and {Shapley}, Alice E. and {Pahl}, Anthony J. and {Strom}, Allison L. and {Lamb}, Noah R. and {Li}, Zhihui and {Rudie}, Gwen C.},
        title = "{The KBSS-KCWI survey: the connection between extended Ly {\ensuremath{\alpha}} haloes and galaxy azimuthal angle at z   2-3}",
      journal = {\mnras},
         year = 2021,
        month = nov,
       volume = {508},
       number = {1},
        pages = {19-43},
          doi = {10.1093/mnras/stab2383},
archivePrefix = {arXiv},
       eprint = {2104.10173},
 primaryClass = {astro-ph.GA},
       adsurl = {https://ui.adsabs.harvard.edu/abs/2021MNRAS.508...19C}
}

@article{oey2007a,
  title = {The {{Survey}} for {{Ionization}} in {{Neutral Gas Galaxies}}. {{III}}. {{Diffuse}}, {{Warm Ionized Medium}} and {{Escape}} of {{Ionizing Radiation}}},
  author = {Oey, M. S. and Meurer, G. R. and Yelda, S. and Furst, E. J. and {Caballero-Nieves}, S. M. and Hanish, D. J. and Levesque, E. M. and Thilker, D. A. and Walth, G. L. and {Bland-Hawthorn}, J. and Dopita, M. A. and Ferguson, H. C. and Heckman, T. M. and Doyle, M. T. and Drinkwater, M. J. and Freeman, K. C. and R. C. Kennicutt, Jr and Kilborn, V. A. and Knezek, P. M. and Koribalski, B. and Meyer, M. and Putman, M. E. and {Ryan-Weber}, E. V. and Smith, R. C. and {Staveley-Smith}, L. and Webster, R. L. and Werk, J. and Zwaan, M. A.},
  year = 2007,
  month = jun,
  journal = {The Astrophysical Journal},
  volume = {661},
  number = {2},
  pages = {801},
  publisher = {IOP Publishing},
  issn = {0004-637X},
  doi = {10.1086/517867},
  urldate = {2025-09-15},
  langid = {english}
}

@article{calabro2023,
  title = {Near-Infrared Emission Line Diagnostics for {{AGN}} from the Local {{Universe}} to {\emph{z}} {$\sim$} 3},
  author = {Calabr{\`o}, Antonello and Pentericci, Laura and Feltre, Anna and Haro, Pablo Arrabal and Radovich, Mario and Seill{\'e}, Lise-Marie and Oliva, Ernesto and Daddi, Emanuele and Amor{\'i}n, Ricardo and Bagley, Micaela B. and Bisigello, Laura and Buat, V{\'e}ronique and Castellano, Marco and Cleri, Nikko J. and Dickinson, Mark and Fern{\'a}ndez, Vital and Finkelstein, Steven L. and Giavalisco, Mauro and Grazian, Andrea and Hathi, Nimish P. and Hirschmann, Michaela and Juneau, St{\'e}phanie and Kartaltepe, Jeyhan S. and Koekemoer, Anton M. and Lucas, Ray A. and Papovich, Casey and {P{\'e}rez-Gonz{\'a}lez}, Pablo G. and Pirzkal, Nor and Santini, Paola and Trump, Jonathan and De La Vega, Alexander and Wilkins, Stephen M. and Yung, L. Y. Aaron and Cassata, Paolo and Gobat, Raphael A. S. and Mascia, Sara and Napolitano, Lorenzo and Vulcani, Benedetta},
  year = 2023,
  month = nov,
  journal = {Astronomy \& Astrophysics},
  volume = {679},
  pages = {A80},
  issn = {0004-6361, 1432-0746},
  doi = {10.1051/0004-6361/202347190},
  urldate = {2025-05-15},
  copyright = {https://creativecommons.org/licenses/by/4.0},
  langid = {english}
}

@ARTICLE{kewley2019notthereview,
       author = {{Kewley}, Lisa J. and {Nicholls}, David C. and {Sutherland}, Ralph and {Rigby}, Jane R. and {Acharya}, Ayan and {Dopita}, Michael A. and {Bayliss}, Matthew B.},
        title = "{Theoretical ISM Pressure and Electron Density Diagnostics for Local and High-redshift Galaxies}",
      journal = {\apj},
         year = 2019,
        month = jul,
       volume = {880},
       number = {1},
          eid = {16},
        pages = {16},
          doi = {10.3847/1538-4357/ab16ed},
archivePrefix = {arXiv},
       eprint = {1908.05504},
 primaryClass = {astro-ph.GA},
       adsurl = {https://ui.adsabs.harvard.edu/abs/2019ApJ...880...16K}
}

@article{ho2014,
  title = {The {{SAMI Galaxy Survey}}: Shocks and Outflows in a Normal Star-Forming Galaxy},
  shorttitle = {The {{SAMI Galaxy Survey}}},
  author = {Ho, I-Ting and Kewley, Lisa J. and Dopita, Michael A. and Medling, Anne M. and Allen, J. T. and {Bland-Hawthorn}, Joss and Bloom, Jessica V. and Bryant, Julia J. and Croom, Scott M. and Fogarty, L. M. R. and Goodwin, Michael and Green, Andy W. and Konstantopoulos, Iraklis S. and Lawrence, Jon S. and {L{\'o}pez-S{\'a}nchez}, {\'A}. R. and Owers, Matt S. and Richards, Samuel and Sharp, Rob},
  year = 2014,
  month = nov,
  journal = {Monthly Notices of the Royal Astronomical Society},
  volume = {444},
  number = {4},
  pages = {3894--3910},
  issn = {1365-2966, 0035-8711},
  doi = {10.1093/mnras/stu1653},
  urldate = {2026-04-22},
  langid = {english}
}

@article{gray2017,
  title = {The {{Effect}} of {{Turbulence}} on {{Nebular Emission Line Ratios}}},
  author = {Gray, William J. and Scannapieco, Evan},
  date = {2017-11-10},
  year= 2017,
  journaltitle = {The Astrophysical Journal},
  shortjournal = {ApJ},
  volume = {849},
  number = {2},
  pages = {132},
  issn = {0004-637X, 1538-4357},
  doi = {10.3847/1538-4357/aa9121},
  url = {https://iopscience.iop.org/article/10.3847/1538-4357/aa9121},
  urldate = {2026-03-17},
  langid = {english}
}

@ARTICLE{rogers2026,
       author = {{Rogers}, Noah S.~J. and {Strom}, Allison L. and {Rudie}, Gwen C. and {Trainor}, Ryan F. and {von Raesfeld}, Caroline and {Raptis}, Menelaos and {Korhonen Cuestas}, Nathalie A. and {Miller}, Tim B. and {Steidel}, Charles C. and {Maseda}, Michael V. and {Chen}, Yuguang and {Law}, David R.},
        title = "{CECILIA: Gas-phase Physical Conditions and Multielement Chemistry at Cosmic Noon}",
      journal = {\apjl},
         year = 2026,
        month = feb,
       volume = {997},
       number = {2},
          eid = {L44},
        pages = {L44},
          doi = {10.3847/2041-8213/ae31f3},
archivePrefix = {arXiv},
       eprint = {2509.18257},
 primaryClass = {astro-ph.GA},
       adsurl = {https://ui.adsabs.harvard.edu/abs/2026ApJ...997L..44R}
}

@ARTICLE{asplund2021,
       author = {{Asplund}, M. and {Amarsi}, A.~M. and {Grevesse}, N.},
        title = "{The chemical make-up of the Sun: A 2020 vision}",
      journal = {\aap},
         year = 2021,
        month = sep,
       volume = {653},
          eid = {A141},
        pages = {A141},
          doi = {10.1051/0004-6361/202140445},
archivePrefix = {arXiv},
       eprint = {2105.01661},
 primaryClass = {astro-ph.SR},
       adsurl = {https://ui.adsabs.harvard.edu/abs/2021A&A...653A.141A}
}

@article{clarendon2025,
  title = {Neutral Oxygen in z {$\sim$} 2--3 Galaxies: {{Probing}} Galaxy Evolution through Faint Emission-Line Diagnostics},
  author = {Clarendon, Audrey and Strom, Allison and {von Raesfeld}, Caroline},
  year = 2025,
  month = jul,
  journal = {Research Notes of the AAS},
  volume = {9},
  number = {7},
  pages = {206},
  publisher = {The American Astronomical Society},
  doi = {10.3847/2515-5172/adf4c5}
}

@article{byler2017,
   title={Nebular Continuum and Line Emission in Stellar Population Synthesis Models},
   volume={840},
   ISSN={1538-4357},
   url={http://dx.doi.org/10.3847/1538-4357/aa6c66},
   DOI={10.3847/1538-4357/aa6c66},
   number={1},
   journal={The Astrophysical Journal},
   publisher={American Astronomical Society},
   author={Byler, Nell and Dalcanton, Julianne J. and Conroy, Charlie and Johnson, Benjamin D.},
   year={2017},
   month=may, pages={44} }

@article{speagle2020,
   title={dynesty: a dynamic nested sampling package for estimating Bayesian posteriors and evidences},
   volume={493},
   ISSN={1365-2966},
   url={http://dx.doi.org/10.1093/mnras/staa278},
   DOI={10.1093/mnras/staa278},
   number={3},
   journal={Monthly Notices of the Royal Astronomical Society},
   publisher={Oxford University Press (OUP)},
   author={Speagle, Joshua S},
   year={2020},
   month=feb, pages={3132–3158} }

@software{kosposov2022,
  author       = {Sergey Koposov and
                  Josh Speagle and
                  Kyle Barbary and
                  Gregory Ashton and
                  Ed Bennett and
                  Johannes Buchner and
                  Carl Scheffler and
                  Ben Cook and
                  Colm Talbot and
                  James Guillochon and
                  Patricio Cubillos and
                  Andrés Asensio Ramos and
                  Ben Johnson and
                  Dustin Lang and
                  Ilya and
                  Matthieu Dartiailh and
                  Alex Nitz and
                  Andrew McCluskey and
                  Anne Archibald and
                  Christoph Deil and
                  Dan Foreman-Mackey and
                  Danny Goldstein and
                  Erik Tollerud and
                  Joel Leja and
                  Matthew Kirk and
                  Matt Pitkin and
                  Patrick Sheehan and
                  Phillip Cargile and
                  ruskin23 and
                  Ruth Angus},
  title        = {joshspeagle/dynesty: v2.0.3},
  month        = dec,
  year         = 2022,
  publisher    = {Zenodo},
  version      = {v2.0.3},
  doi          = {10.5281/zenodo.7388523},
  url          = {https://doi.org/10.5281/zenodo.7388523},
}

@ARTICLE{li2025-cue,
       author = {{Li}, Yijia and {Leja}, Joel and {Johnson}, Benjamin D. and {Tacchella}, Sandro and {Davies}, Rebecca and {Belli}, Sirio and {Park}, Minjung and {Emami}, Razieh},
        title = "{Cue: A Fast and Flexible Photoionization Emulator for Modeling Nebular Emission Powered by Almost Any Ionizing Source}",
      journal = {\apj},
         year = 2025,
        month = jun,
       volume = {986},
       number = {1},
          eid = {9},
        pages = {9},
          doi = {10.3847/1538-4357/adcab4},
archivePrefix = {arXiv},
       eprint = {2405.04598},
 primaryClass = {astro-ph.GA},
       adsurl = {https://ui.adsabs.harvard.edu/abs/2025ApJ...986....9L}
}

@ARTICLE{stanway2018bpassv2.2.1,
       author = {{Stanway}, E.~R. and {Eldridge}, J.~J.},
        title = "{Re-evaluating old stellar populations}",
      journal = {\mnras},
         year = 2018,
        month = sep,
       volume = {479},
       number = {1},
        pages = {75-93},
          doi = {10.1093/mnras/sty1353},
archivePrefix = {arXiv},
       eprint = {1805.08784},
 primaryClass = {astro-ph.GA},
       adsurl = {https://ui.adsabs.harvard.edu/abs/2018MNRAS.479...75S}
}

@misc{cloudy23,
      title={The 23.01 release of Cloudy}, 
      author={Chamani M. Gunasekera and Peter A. M. van Hoof and Marios Chatzikos and Gary J. Ferland},
      year={2023},
      eprint={2311.10163},
      archivePrefix={arXiv},
      primaryClass={astro-ph.GA},
      url={https://arxiv.org/abs/2311.10163}, 
}

@article{allen2008,
   title={The MAPPINGS III Library of Fast Radiative Shock Models},
   volume={178},
   ISSN={1538-4365},
   url={http://dx.doi.org/10.1086/589652},
   DOI={10.1086/589652},
   number={1},
   journal={The Astrophysical Journal Supplement Series},
   publisher={American Astronomical Society},
   author={Allen, Mark G. and Groves, Brent A. and Dopita, Michael A. and Sutherland, Ralph S. and Kewley, Lisa J.},
   year={2008},
   month=sep, pages={20–55} }

@article{dopita2000,
  title = {A {{Theoretical Recalibration}} of the {{Extragalactic HII Region Sequence}}},
  author = {Dopita, M. A. and Kewley, L. J. and Heisler, C. A. and Sutherland, R. S.},
  year = {2000},
  month = oct,
  journal = {The Astrophysical Journal},
  volume = {542},
  number = {1},
  pages = {224},
  publisher = {IOP Publishing},
  issn = {0004-637X},
  doi = {10.1086/309538},
  urldate = {2025-09-02},
  langid = {english}
}

@ARTICLE{korhonencuestas2025,
       author = {{Korhonen Cuestas}, Nathalie A. and {Strom}, Allison L. and {Miller}, Tim B. and {Steidel}, Charles C. and {Trainor}, Ryan F. and {Rudie}, Gwen C. and {Nu{\~n}ez}, Evan Haze},
        title = "{Exploring the Relationship between Stellar Mass, Metallicity, and Star Formation Rate at z {\ensuremath{\sim}} 2.3 in KBSS-MOSFIRE}",
      journal = {\apj},
         year = 2025,
        month = may,
       volume = {984},
       number = {2},
          eid = {188},
        pages = {188},
          doi = {10.3847/1538-4357/adc5f7},
archivePrefix = {arXiv},
       eprint = {2503.10800},
 primaryClass = {astro-ph.GA},
       adsurl = {https://ui.adsabs.harvard.edu/abs/2025ApJ...984..188K}
}

@ARTICLE{brinchmann2023,
       author = {{Brinchmann}, Jarle},
        title = "{High-z galaxies with JWST and local analogues - it is not only star formation}",
      journal = {\mnras},
         year = 2023,
        month = oct,
       volume = {525},
       number = {2},
        pages = {2087-2106},
          doi = {10.1093/mnras/stad1704},
archivePrefix = {arXiv},
       eprint = {2208.07467},
 primaryClass = {astro-ph.GA},
       adsurl = {https://ui.adsabs.harvard.edu/abs/2023MNRAS.525.2087B}
}

@ARTICLE{shapley2019,
       author = {{Shapley}, Alice E. and {Sanders}, Ryan L. and {Shao}, Peng and {Reddy}, Naveen A. and {Kriek}, Mariska and {Coil}, Alison L. and {Mobasher}, Bahram and {Siana}, Brian and {Shivaei}, Irene and {Freeman}, William R. and {Azadi}, Mojegan and {Price}, Sedona H. and {Leung}, Gene C.~K. and {Fetherolf}, Tara and {de Groot}, Laura and {Zick}, Tom and {Fornasini}, Francesca M. and {Barro}, Guillermo},
        title = "{The MOSDEF Survey: Sulfur Emission-line Ratios Provide New Insights into Evolving Interstellar Medium Conditions at High Redshift}",
      journal = {\apjl},
         year = 2019,
        month = aug,
       volume = {881},
       number = {2},
          eid = {L35},
        pages = {L35},
          doi = {10.3847/2041-8213/ab385a},
archivePrefix = {arXiv},
       eprint = {1907.07189},
 primaryClass = {astro-ph.GA},
       adsurl = {https://ui.adsabs.harvard.edu/abs/2019ApJ...881L..35S}
}

@ARTICLE{mendez-delgado2023,
       author = {{M{\'e}ndez-Delgado}, J.~E. and {Esteban}, C. and {Garc{\'\i}a-Rojas}, J. and {Arellano-C{\'o}rdova}, K.~Z. and {Kreckel}, K. and {G{\'o}mez-Llanos}, V. and {Egorov}, O.~V. and {Peimbert}, M. and {Orte-Garc{\'\i}a}, M.},
        title = "{Density biases and temperature relations for DESIRED H II regions}",
      journal = {\mnras},
         year = 2023,
        month = aug,
       volume = {523},
       number = {2},
        pages = {2952-2973},
          doi = {10.1093/mnras/stad1569},
archivePrefix = {arXiv},
       eprint = {2305.13136},
 primaryClass = {astro-ph.GA},
       adsurl = {https://ui.adsabs.harvard.edu/abs/2023MNRAS.523.2952M}
}

@BOOK{osterbrock2006,
       author = {{Osterbrock}, Donald E. and {Ferland}, Gary J.},
        title = "{Astrophysics of gaseous nebulae and active galactic nuclei}",
         year = 2006,
        publisher = {University Science Books},
       adsurl = {https://ui.adsabs.harvard.edu/abs/2006agna.book.....O}
}
\bibliographystyle{aasjournalv7.1}

\end{CJK*}
\end{document}